\documentclass[letterpaper]{JHEP3}
\usepackage{epsfig,axodraw,graphicx,psfrag,amsmath,subfigure} 

\newcommand{\slal}{l\!\!/}
\newcommand{\slap}{p\!\!\!/}
\newcommand{\slapp}{p^\prime \!\!\!\!/}
\newcommand{\met}{\rlap{\,/}E_T}
\newcommand{\one}{{({\bf 1})} }
\newcommand{\e}{\mu_\circ}
\newcommand{\mup}{\mu_{\star}}

\newcommand{\be}{\begin{equation}}
\newcommand{\ee}{\end{equation}}
\newcommand{\bear}{\begin{eqnarray}}
\newcommand{\eear}{\end{eqnarray}} 
\newcommand{\ba}{\begin{array}}
\newcommand{\ea}{\end{array}}
\newcommand{\lae}{\begin{array}{c}\,\sim\vspace{-21pt}\\<
\end{array}}

\title{Leptons and photons at the LHC: cascades through spinless adjoints
\\ { $\; $ } \\ }

\author{Bogdan A.~Dobrescu, Kyoungchul Kong, Rakhi Mahbubani \\ \\ 
Theoretical Physics Department, Fermilab, Batavia, IL 60510, USA \\ \\
\email{bdob@fnal.gov, kckong@fnal.gov, rakhi@fnal.gov} \\ 
}

\abstract{ 
We study the hadron collider phenomenology of (1,0)
Kaluza-Klein modes along two universal extra dimensions 
compactified on the chiral square. 
Cascade decays of spinless adjoints proceed 
through  tree-level 3-body decays involving leptons as well as
one-loop 2-body decays involving photons.
As a result, spectacular events with as many as six
charged leptons, or one photon plus four charged leptons are 
expected to be observed at the LHC.
Unusual events with relatively large branching fractions include three 
leptons of same charge plus one lepton of opposite charge, 
or one photon plus two leptons of same charge.
We estimate the current limit from the Tevatron on the compactification 
scale, set by searches for trilepton events, to be around 270 GeV.
  \\ \\ }

\preprint{hep-ph/0703231 \\ {\small FERMILAB-PUB-07-062-T} \\ March 21, 2007 \\ } 

\begin{document}


\bigskip

\section{Introduction} \setcounter{equation}{0}

Theories beyond the standard model which include 
several new particles at the TeV scale and a new 
discrete symmetry lead to cascade decays with interesting 
signatures at colliders. At the same time, the discrete symmetry
reduces the contributions of new particles to electroweak 
observables, allowing the new particles to be light enough such that 
they can be copiously produced not only at the LHC, but perhaps even at the
Tevatron.  Classic examples of such theories
include supersymmetric models with $R$-parity, 
universal extra dimensions \cite{Appelquist:2000nn},
and Little Higgs models with  $T$-parity \cite{Cheng:2003ju}.
Typically, the cascade decays in these models lead to observable 
events with up to four leptons and missing transverse energy 
\cite{Cheng:2002ab, Datta:2005zs}. 

In this paper we show that more spectacular events,
with five or six leptons, or one photon and several leptons,
are predicted in the 6-dimensional standard model (6DSM).
This model  \cite{Burdman:2006gy}, in which all 
standard model particles propagate 
in two universal extra dimensions compactified on the 
chiral square \cite{Burdman:2005sr,Dobrescu:2004zi,Hashimoto:2004xz}, is motivated 
by the prediction based on anomaly cancellations 
that the number of fermion generations is a multiple of three
\cite{Dobrescu:2001ae}, 
and by the long proton lifetime enforced by a remnant of
6D Lorentz symmetry \cite{Appelquist:2001mj}.

The  larger number of leptons and the presence of photons is due
to the existence of `spinless adjoint' particles, the 
Kaluza-Klein (KK) modes of gauge bosons polarized 
along extra dimensions. Compared to five-dimensional (5D) models
where such fields become the longitudinal components of the
KK vector bosons, in six-dimensional (6D) gauge theories there is 
an additional field for each KK vector boson, which represents
a physical spin-0 particle transforming in the adjoint 
representation of the gauge group \cite{Burdman:2006gy}. 

The 6DSM has a KK parity corresponding to reflections with respect to
the center of the chiral square. Its consequences are similar to 
the ones in the case of a single universal extra dimension \cite{Hooper:2007qk}, 
where KK parity is the symmetry under reflections with respect to the center of
the compact dimension. It is well known that in the 5D case KK parity 
ensures the stability of the lightest KK particle (LKP). Furthermore, 
loop corrections select the KK mode of the hypercharge boson to be the 
LKP \cite{Cheng:2002iz}, and that is a viable dark matter candidate 
\cite{Servant:2002aq}.
The same is true in the 6DSM, with the additional 
twist that the LKP in that case is a spinless adjoint.
In fact, 
one-loop mass corrections in this model lift the degeneracy of the
modes at each KK level, making all spinless adjoints lighter than 
the corresponding
vector bosons \cite{Ponton:2005kx}.

Particles on the first KK level, having KK numbers (1,0), 
are odd under  KK parity. As a result, they may be produced only in pairs
at colliders, 
and each of their cascade decays produces an LKP, which is seen as 
missing transverse energy in the detector.
The goal of this paper is to determine the main 
signatures of (1,0) particles at hadron colliders. 
Particles on the second level, which have KK numbers (1,1) 
and are even under KK parity, lead to a completely different set of 
signatures, mainly involving resonances of top and bottom quarks 
\cite{Burdman:2006gy}. 

We review the 6DSM in
Section \ref{sec:couplings}, and then proceed in Section \ref{sec:decays} to
calculate decay widths for $(1,0)$ modes.  We  analyze the 
production of these particles at the LHC and Tevatron in Section
\ref{sec:production}, and compute rates for events with leptons 
and photons.
Several comments regarding our results are given in Section \ref{sec:conclusions}.  
Feynman rules for this
model are given in Appendix A. Details of the calculations
of one-loop 2-body and tree-level 3-body decay widths for
spinless adjoints and vector bosons can be found in 
Appendices B and C, respectively.

\bigskip

\section{Two universal extra dimensions}\label{sec:model}
\label{sec:couplings}
\setcounter{equation}{0}

We assume that all standard model fields propagate in two flat extra
dimensions, of coordinates $x_4$ and $x_5$, compactified on a square
of side $L=\pi R$ with adjacent sides identified in pairs (see Figure 1). This
compactification predicts that the fermion zero modes are
chiral, and therefore may represent the observed quarks and
leptons. Furthermore, this `chiral square' is invariant under rotations by
$\pi$ about its center.  The ensuing $Z_2$ symmetry,
known as KK parity, implies that the lightest KK-odd particle is stable.

\begin{figure}[t]
\vspace*{-4mm}
    \centering    
\includegraphics[width=.45 \textwidth]{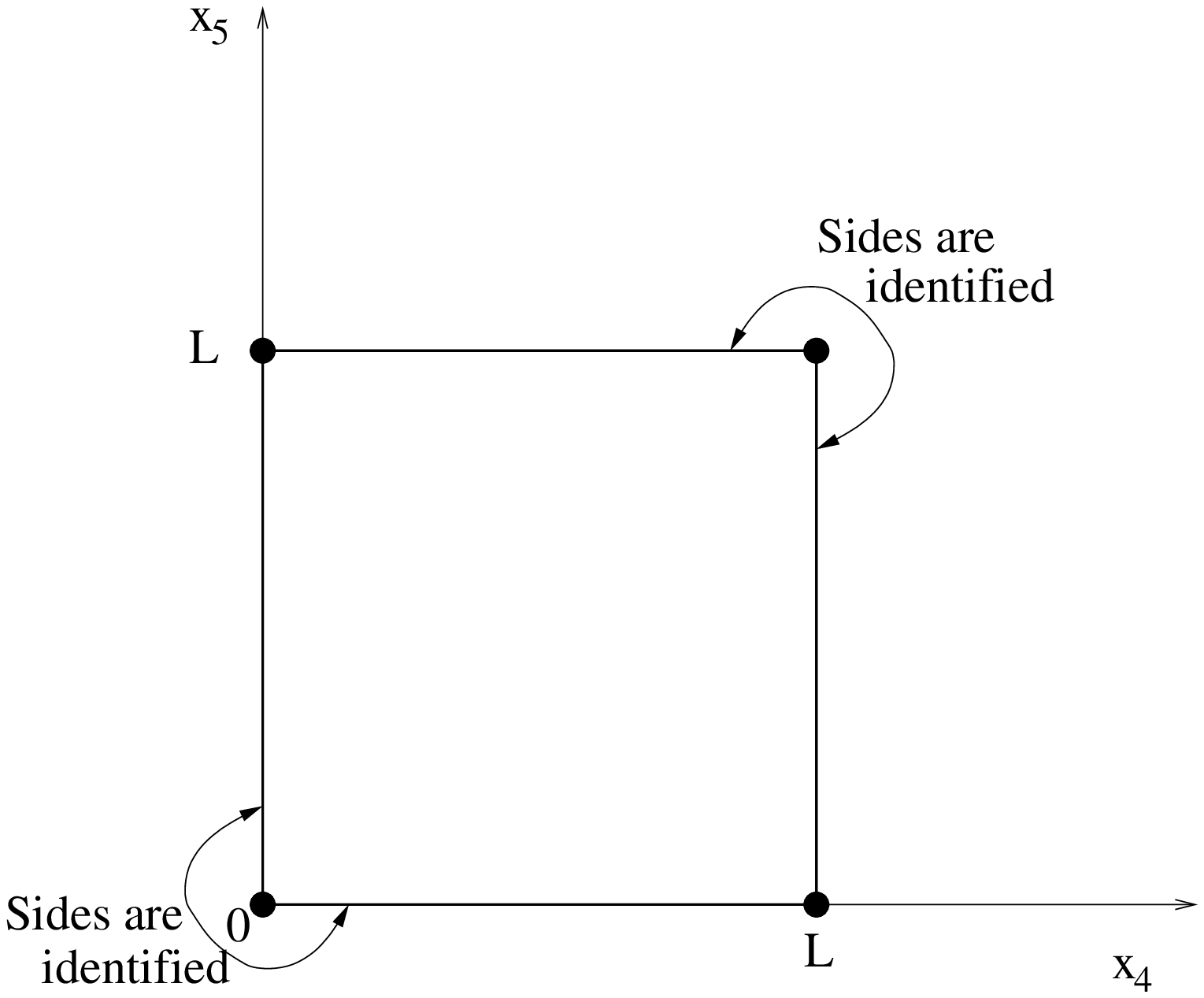}
\includegraphics[width=.45 \textwidth]{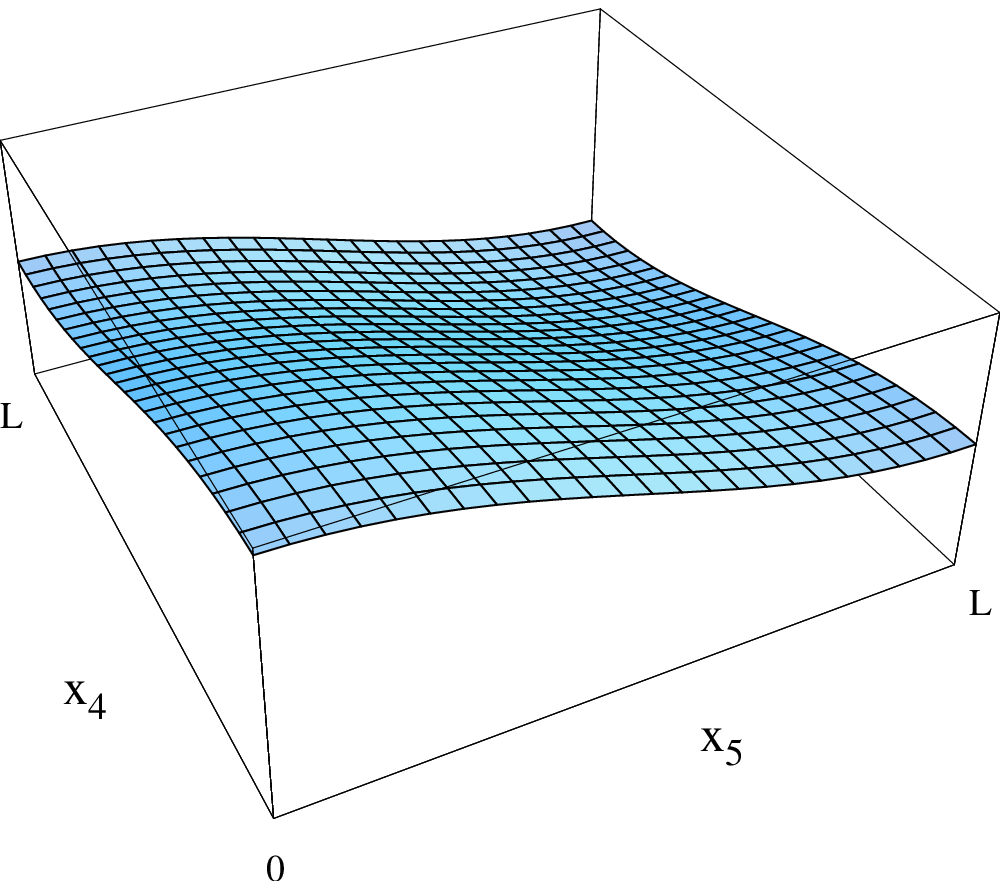}
\caption{Chiral square compactification (left) and level-1 KK function $f_0^\one (x_4,x^5)$ 
for standard model fields (right).}\bigskip   
\label{fig:fprofile}
\end{figure}

Equality of the Lagrangian densities on adjacent sides of the square is
achieved by enforcing that bulk fields and their first derivatives
vary smoothly across the boundary.  Applying these boundary
conditions to solve the 6D equations of motion for these fields,
by separation of variables, we
find that the dependence on $x_4$ and $x_5$ can be expressed in terms of one of
four complete and orthonormal sets of functions $f_n^{(j,k)}$ with
$n=0,1,2,3$, where
the KK numbers $(j,k)$ are integers and $j\geq 1$, $k\geq 0$ or
$j=k=0$. All $(j,k)$ modes have tree-level mass $\sqrt{j^2+k^2}/R$
before electroweak symmetry breaking.  

\bigskip \bigskip

\subsection{Interactions of the (1,0) modes}\label{sec:interactions}

We are
primarily interested in the phenomenology of the $(1,0)$ modes here.
We loosely refer to these as `level-1' modes because they are the
lightest nonzero KK modes.  For notational brevity we will label
them using the superscript $\one$.

The level-1 KK modes belonging to a tower that includes a zero mode
has a KK function
\be
f^\one_{0}(x_4,x_5)=\cos{\left(\frac{x_4}{R}\right)}+\cos{\left(\frac{x_5}{R}\right)}~,
\ee
which is plotted in Figure \ref{fig:fprofile}.
This is the case for the KK modes of all spin-1 fields and fermions of the
same chirality as the observed quarks and leptons, as well as the
Higgs doublet.  The spinless adjoint field, $A^\one_H$, which is the uneaten
combination of the extra-dimensional polarizations of the 6D gauge field,
is associated with a KK function which is independent of $x_4$,
\be
f^\one_H= - \frac{1}{2}\left[ f^\one_1(x_4,x_5)-f^\one_3(x_4,x_5) \right]
= -\sin{\frac{x_5}{R}}~,
\ee
while the longitudinal component of the vector KK modes is 
associated with a KK function which is independent of $x_5$:
\be
f^\one_G= - \frac{i}{2}\left[f^\one_1(x_4,x_5)+f^\one_3(x_4,x_5) \right]
= \sin{\frac{x_4}{R}}~.
\ee
KK modes of fermions come in vectorlike pairs with the component of 
4D chirality opposite to the corresponding standard model fermion having KK
function $f_1$ or $f_3$,
depending on the 6D chirality.

Integrating
over the extra dimensional coordinates gives
the 4D effective Lagrangian, which contains kinetic and interaction 
terms for all SM particles and their KK modes.  We limit
ourselves to detailing in this section only the couplings
of the standard model fields with the level-1 KK modes; the latter are odd
under KK parity and so only appear in pairs.
The general Lagrangian
for all modes is derived in Ref.~\cite{Burdman:2005sr,Dobrescu:2004zi}, while the
couplings for all fermion modes can be found in Appendix B.

The $SU(3)_c$ gauge interactions
include the following tree-level couplings between zero modes and $(1,0)$
modes:
%
\bear
\mathcal{L}_{\rm gauge} &\supset & 
g_sf^{abc}\left[ G_\mu^{\one a}\left(\partial^\mu G^{\nu\one b}
- \partial^\nu G^{\mu\one b}\right) G_\nu^{\one c}
- G_\mu^{\one a}G_\nu^{\one b}\partial^\mu G^{\nu c}
+ G_H^{\one a}\partial^\mu G_H^{\one b} G_\mu^{c}\right]
\nonumber\\ [2mm]
&-& \frac{g_s^2}{2}\left[f^{abd}f^{ace}G_\mu^{\one b}G^{\mu\one c}G_\nu^dG^{\nu e}
+ \left(f^{abc}f^{ade}+f^{adc}f^{abe}\right)
G_\mu^{\one b}G^{\mu d}G_\nu^{\one c}G^{\nu e}\right] 
\nonumber\\ [2mm]
&+& \frac{g_s^2}{2}f^{abc}f^{ade}G_H^{\one c}G_H^{\one e}G_\mu^{b} G^{\mu d} ~,
\eear
where $g_s$ is the
QCD gauge coupling, $f^{abc}$ are the $SU(3)_c$ structure constants, 
and $G_{\mu}^\one$ and $G_H^\one$ are the level-1
vector and spinless adjoint KK modes of the gluon $G_\mu$. We have
suppressed all superscripts for zero modes.
There are also interactions of the quark modes with the QCD vector and
spinless modes:
\be
\mathcal{L}_{\rm matter}\supset\!\!\!\displaystyle\sum_{\rm fermions}\!\!\!g_s
\overline{Q}_{\pm}^\one G_\mu^a T^a\gamma^\mu Q^\one_{\pm}+g_s\bigg[
\overline{Q}_{\pm}^\one G_\mu^{\one a}T^a\gamma^\mu P_{\substack{L\\R}}Q_{\pm}
- i\overline{Q}_{\pm}^\one G_H^{\one a}T^a P_{\substack{L\\R}}Q_{\pm}
+ {\rm H.c.}\bigg] ~,
\ee
where fermions with 6D chirality $+$ contain left-handed zero modes, and
fermions with 6D chirality  $-$ contain right-handed zero modes.  The
$SU(2)_W$ and $U(1)_Y$ sectors are
analogous, with all the gauge self-couplings set to zero in the Abelian
case.  The Higgs and ghost terms are given in Ref.~\cite{Burdman:2005sr,Dobrescu:2004zi}.

\bigskip

\subsection{Mass corrections}\label{sec:masscorrections}


Computing radiative corrections in this theory involves taking sums over KK modes, or momenta in the
extra dimensions, which fourier transform to operators localized at
the corners of the chiral square, $(0,0)$, $(\pi R,\pi R)$ and $(0,\pi
R)\sim(\pi R,0)$.  The most
general 4D effective Lagrangian must therefore allow for these ~\cite{Ponton:2005kx}:
\be
L_{eff}=\int_0^L d x^4\int_0^L d x^5
\left[\mathcal{L}_{\rm bulk}+\bigg(\delta(x_4)\delta(x_5)+\delta(L-x_4)\delta(L-x_5) 
\bigg)\mathcal{L}_1+\delta(L-x_5)\mathcal{L}_2\right]~,
\ee
where $\mathcal{L}_1$ and $\mathcal{L}_2$ contain all localized operators.  Note that 
KK parity ensures the equality of the operators localized  
at $(0,0)$ and $(L,L)$.  Local operators 
break 6D Lorentz invariance and hence give rise to mass corrections
for KK particles.   Such terms are important for models of
flat extra dimensions since they allow for the decays of higher modes
into pairs of lower ones, a process which would otherwise be on
threshold at best due to the quantization of KK mode masses. They 
also make for a more interesting phenomenology by lifting the 
degeneracy of states at each level.

The localized terms contain contributions from ultraviolet physics 
as well as from running down from the cut-off.  Being unable to
compute the former, we assume that they are generically smaller than the
logarithmically-enhanced one-loop terms which are calculable (for further 
discussion see \cite{Ponton:2005kx,Cheng:2002iz}).  Level-1 fermions 
acquire the following mass corrections ~\cite{Burdman:2006gy}:
\bear
\delta(M_{Q_+}) &=& \left(\frac{16}{3}g_s^2+3g^2+\frac{1}{9}g'^2
+ \frac{5}{8}\lambda_{Q_+}^2\right)\frac{l_0}{R}
+ \frac{1}{2}m_{q}^2 R~,
\nonumber\\  [2mm] 
\delta(M_{Q_-}) &=& 
\left(\frac{16}{3}g_s^2+4g'^2 y^2+\frac{10}{8}\lambda_{Q_-}^2\right)\frac{l_0}{R}
+ \frac{1}{2}m_{q}^2 R ~, 
\nonumber\\ [2mm] 
\delta(M_{L_+})&=&\left(3g^2+g'^2\right)\frac{l_0}{R}~,
\nonumber\\ [2mm] 
\delta(M_{E_-})&=&\frac{g'^2}{4\pi^2}\frac{l_0}{R} ~,
\label{fermion-corr}
\eear
where  $g_s$, $g$ and $g^\prime$ are the $SU(3)_c\times SU(2)_W\times U(1)_Y$
gauge couplings, $\lambda_{Q_\pm}$ are the Yukawa couplings of $Q_\pm$ to the Higgs
doublet, and $l_0$ is a common loop factor,
\be
l_0=\frac{1}{16\pi^2}\ln \left(\Lambda R\right)^2  ~.
\ee 
An estimate of the cutoff of the effective theory, based on naive dimensional analysis,
 gives $\Lambda \approx 10/R$ 
\cite{Burdman:2006gy}. The terms linear in $R$ shown in 
Eq.~(\ref{fermion-corr})
are small corrections to the tree-level masses due to electroweak symmetry breaking masses, $m_q$.  

The (1,0) vector bosons also receive radiative corrections to their masses,
\bear
\delta M_{G_\mu^\one} &=& 4 g_s^2\frac{l_0}{R} ~, 
\nonumber\\ [2mm]
\delta M_{W_\mu^\one} &=&\frac{123}{24}g^2 \frac{l_0}{R}  ~,
\nonumber\\ [2mm]
\delta M_{B_\mu^\one} &=&-\frac{165}{24}g'^2 \frac{l_0}{R}  ~,
\eear
while only the spinless adjoints in the electroweak sector have mass corrections:
\bear
\delta M_{G_H^\one} &=& 0 
\nonumber\\ [2mm]
\delta M_{W_H^\one} &=& -\frac{51}{8}g^2 \frac{l_0}{R} +\frac{m_{W}^2 R}{2}  ~, 
\nonumber\\ [2mm]
\delta M_{B_H^\one} &=& -\frac{307}{8}g'^2 \frac{l_0}{R}  ~.
\eear
The above mass shifts include negative contributions from fermions
in loops, allowing for overall negative corrections to masses.  
This is especially important when there are no self-interactions to compete
with the fermion interactions, as is the case with for the hypercharge bosons.

\begin{table}[t]
\hspace*{-.2cm}
\renewcommand{\arraystretch}{1.7}
\begin{tabular}{|c|c||c|c|}
\hline \ \hspace*{-.28cm}boson\hspace*{-.28cm} \ & $M R$ \ 
& \hspace*{-.16cm}fermion\hspace*{-.17cm} & $M R$ \
\rule{0mm}{5mm}\rule{0mm}{-22mm} \\ \hline\hline 
$G_\mu^\one$ & 1.392  & $Q_+^{\one 3}$ & 
$\!1.265 + \frac{1}{2}(m_t R)^2\!$  \\ \hline
$\! W_\mu^\one$ & $1.063  + \frac{1}{2}(M_W R)^2\!$ & $T_-^\one$ & 
$\! 1.252 +  \frac{1}{2}(m_t R)^2\!$  \\ \hline
$G_H^\one$   & 1.0   & $Q_+^\one$ & 1.247 \\ \hline 
$B_\mu^\one$ & 0.974 & $U_-^\one$ & 1.216 \\ \hline
$\! W_H^\one$   & $0.921 + \frac{1}{2}(m_W R)^2\!$  & $D_-^\one$ & 1.211 \\ \hline
$B_H^\one$   & 0.855 & $L_+^\one$ & 1.041  \\ \hline
\multicolumn{2}{r|}{} & $E_-^\one$ & 1.015 \\ \cline{3-4}
\end{tabular} 
\psfrag{mass}[B]{\hspace*{3.5em} $M$ [GeV] }  \small
\psfrag{gmu}[B]{\hspace*{6.5mm} $G^\one_\mu$}
\psfrag{wmu}[B]{\hspace*{6.5mm} $W^\one_\mu$}
\psfrag{bmu}[B]{\hspace*{6.5mm} $B^\one_\mu$}
\psfrag{gh}[B]{\hspace*{-2.5mm} $G^\one_H$}
\psfrag{wh}[B]{\hspace*{-2.5mm} $W^\one_H$}
\psfrag{bh}[B]{\hspace*{-2.5mm} $B^\one_H$}
\psfrag{qp3}[B]{\hspace*{.5mm} $Q_+^{3 \one}$}
\psfrag{qp}[B]{\hspace*{.5mm} $Q_+^\one$}
\psfrag{dm}[B]{\hspace*{.5mm} $D_-^\one$}
\psfrag{tm}[B]{\hspace*{-2.5mm} $T_-^{\one}$}
\psfrag{um}[B]{\hspace*{-2.1mm} $U_-^{\one}$}
\psfrag{lp}[B]{\hspace*{-2.5mm} $L_+^{\one}$}
\psfrag{em}[B]{\hspace*{-2.5mm} $E_-^{\one}$}
\psfrag{RR}[T]{\hspace*{-1.9cm} $1/R = 500$ GeV}

\vspace*{-9.cm}
\hspace*{9.4cm}
\psfig{ file=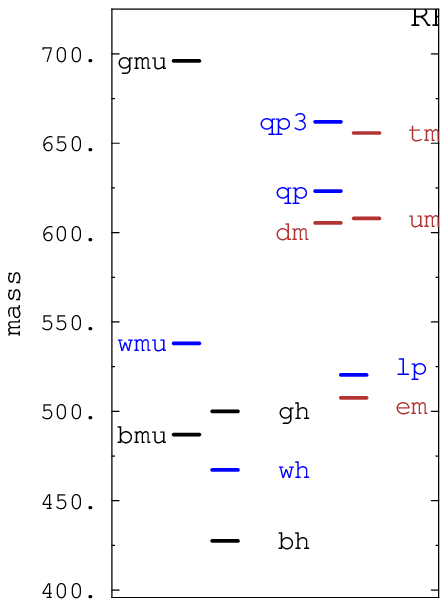,width=6.1cm,angle=0}
\vspace*{-1.9cm}
\caption{Masses of the (1,0) particles in $1/R$ units (left). The (1,0) Higgs particles are 
not included here because their masses are quadratically sensitive to the cutoff scale.
The right-hand panel shows the 
spectrum for $1/R = 0.5$ TeV.}
\label{tab:mass} 
\end{table}

%
%

The masses of the (1,0) particles are given in Table 1 in units of $1/R$.
The mass shifts are evaluated there for gauge couplings $g_s=1.16$, $g=0.65$ and
$g'=0.36$, which are the values obtained using the 
standard model one-loop running up to the scale $1/R=500$ GeV,
We will use the masses from  Table 1 throughout the paper, ignoring 
further running of the gauge couplings above 500 GeV (note that the standard 
model running of the gauge couplings between 500 GeV and 1 TeV results in 
only a 3\% change in $g_s$ and negligible changes in $g$ and $g^\prime$;
however, above $\sim 1/R$ the running is accelerated by the presence of 
the level-1 modes). 

The KK modes of the Higgs doublet have mass-squared shifts which are quadratically
sensitive to the cutoff scale $\Lambda$ \cite{Cheng:2002iz}. Hence, the masses
of the (1,0) Higgs scalars may be treated as free parameters (determined by the 
underlying theory above $\Lambda$, which is not specified in our framework).
Furthermore, 
additional structures such as the Twin Higgs mechanism \cite{Chacko:2005pe} 
may be used to cancel the quadratic divergences in models with universal extra dimensions
\cite{Burdman:2006jj}, potentially affecting the (1,0) Higgs sector. 
We assume here that the (1,0) Higgs particles are heavier than $1/R$.
In that case, the hadron collider phenomenology is mostly independent of the exact 
(1,0) Higgs masses.

\bigskip

\subsection{Loop-induced bosonic operators}

In addition to lifting the degeneracy of the $(1,0)$ masses, loop corrections also contribute to the following 
dimension-5 operators that are of particular interest for computing the branching fractions of the $(1,0)$ bosons:
\be
-\frac{R}{4} \Big ( \mathcal{C}_B \epsilon^{\mu\nu\alpha\beta} F_{\mu\nu} B^\one_{\alpha\beta} B_H^\one 
+ \mathcal{C}_G \epsilon^{\mu\nu\alpha\beta}  G_{\mu\nu} B^\one_{\alpha\beta} G_H^\one \Big )
~,
\label{operator}
\ee
where $F_{\mu\nu}$ and $G_{\mu\nu}$ are the field strengths of the photon and gluon, respectively,
$B^\one_{\alpha\beta}$ is the field strength 
of the $(1,0)$ hypercharge vector boson $B^\one_\alpha$, and $B_H^\one$ is the 
$U(1)_Y$ spinless adjoint.  These operators account for the 
only significant 2-body decay channels open to the
level-1 KK modes $G^\one_H$ and $B^\one_\mu$.
The analogous operator with the photon replaced by the $Z$ boson is less relevant
because the corresponding decay width is phase-space suppressed.
The coefficients of the above dimension-5 operators are computed in
Appendix B, with the result:
\be
\mathcal{C}_B = \frac{g^{\prime 2} e}{8 \pi^2 R} 
\frac{1}{M_{B_\nu^\one }^2 - M_{B_H^\one }^2} \sum_F  \sigma_F \Big ( \frac{Y_F}{2} \Big )^2 Q_F {\cal E}_F
~,
\label{operator-coef}
\ee
where $\sigma_F = \pm 1$ for a 6D fermion $F$ of chirality $\pm$, $Q_F$ is the electric charge,
$Y_F$ is the hypercharge normalized 
to be twice the electric charge for $SU(2)_W$ singlets
and ${\cal E}_F$ is a function of the masses of $B_H^\one$, $B_\nu^\one$,
and of the (1,0) and (1,1) fermions given in Eq.~(\ref{ef}).  $\mathcal{C}_G$ is given
by an analogous expression, but it is suppressed by the small mass difference between the initial-
and final-state $(1,0)$ bosons.

One might also naively expect higher-dimension operators of the form 
\be
G_{\mu\nu}\partial^\mu B_H^\one\partial^\nu G_H^\one
+ Z_{\mu\nu}\partial^\mu B_H^\one\partial^\nu W_H^{\one 3}
+ \left(W_{\mu\nu}^+\partial^\mu B_H^\one\partial^\nu W_H^{\one -} + {\rm H.c.}\right)
~,
\ee
to be generated, where $W^\one_H$ is the level-1 $SU(2)_W$ spinless adjoint and $W_{\mu\nu}$ 
and $Z_{\mu\nu}$ are the standard 
model field strengths for the $W$ and $Z$ bosons.  However, the first of these terms is 
identically zero as can be seen after integrating by parts and using 
the gluon field equation. By the same method one can see that the coefficients of the last two terms are 
small, being proportional to $(m_W R)^2$, and furthermore the resulting decay widths for $W_H^{\one}$ are 
also phase-space suppressed. 

\bigskip\bigskip

\section{Decays of the level-1 particles}
\label{sec:decays}
\setcounter{equation}{0}

KK parity allows any (1,0) particle to decay only into a lighter (1,0)
particle and one or more standard model particles. The lightest
(1,0) particle is stable. In this section we compute the branching fractions
of the (1,0) particles assuming that the generic features of the `one-loop' 
mass spectrum, shown in Table~\ref{tab:mass}, are not modified by higher-order 
corrections.

\subsection{Color-singlet $(1,0)$ particles}

The $W^{\one }_H$ boson (the spinless adjoint of $SU(2)_W$) is the next-to-lightest (1,0)
particle, and therefore can decay only into a $B^{\one }_H$
plus standard model particles. The dominant decay mode
of its electrically neutral component is the 3-body decay 
$W^{\one 3}_H \to B^{\one }_{H} l\bar{l}$, where 
$l$ are leptons. The width for this decay, computed in 
Appendix~C,
is given by 
\be
\Gamma\left(W^{\one 3}_H \to B^{\one }_{H} e^+e^-\right) 
= \, 
\frac{\alpha^{2}  \, M_{W_H^{\one }} }{128\pi\cos^2\!\theta_w\,\sin^2\!\theta_w }
\, {\cal I}_+\!\left( M_{W_H^\one}, M_{B_H^\one}, M_{L_+^\one}\right)  ~,
\ee
and is the same for any lepton pair. The dimensionless function ${\cal I}_+$ contains 
phase space integrals for the decay and is defined in Eq.~(\ref{fpm}).
Expanding this to leading order in the mass difference $M_{W_H^\one} - M_{B_H^\one}$,
which  is accurate to about 25\% for the mass spectrum in Table~\ref{tab:mass} 
[see Eq.~(\ref{fpe}) in Appendix~C], we find that 
the width of the $W^{\one 3}_H$ decay into $B^{\one }_{H}$ plus quarks
has a simple expression in terms of the decay width into
$B^{\one }_{H}$ plus leptons:
\be
\Gamma\left(W^{\one 3}_H \to B^{\one }_{H} q\overline{q} \right) 
\approx \frac{1}{3} \left( \frac{ M_{L_+^{\one }}^2 - M^2_{W_H^{\one }} }{
 M_{Q_+^{\one }}^2 - M^2_{W_H^{\one }} } \right)^{\!\! 4}  
\Gamma\left(W^{\one }_H \to B^{\one }_{H} e^+e^-\right) ~,
\ee
where we have not summed over quark flavors.
Given that $W_H^{\one }$ is closer to $L_+^{\one }$ in mass than to $Q_+^{\one }$, 
it follows that the decay into quarks is highly suppressed. 
The ensuing branching fractions for the $W_H^{\one \, 3} \to B^{\one }_{H}$
transition are approximately 1/6 for each of the $e^+e^-$, $\mu^+\mu^-$
and $\tau^+\tau^-$ final states, 1/2 for $\nu\overline{\nu}$, and 
$0.5$\% for the sum of all quark-antiquark pairs.

The electrically charged spinless adjoints of $SU(2)_W$, $W_H^{\one \, \pm}$,
decay with a branching fraction of nearly 1/3 into each of the 
$e^\pm\nu B^{\one }_{H}$, $\mu^\pm\nu B^{\one }_{H}$ and 
$\tau^\pm\nu B^{\one }_{H}$ final states, while the branching fraction into 
$q\overline{q} B^{\one }_{H}$ is again negligible.

The spin-1 boson $B^{\one }_\mu$ may decay only into a $B^{\one }_H$ or $W^{\one }_H$  
and standard model particles. 
An important tree-level decay is into right-handed leptons and a $B^{\one }_H$, with 
a width:
\be
\Gamma\left(B^\one_\mu \to B^{\one }_{H} e_R^+e_R^-\right) 
= \frac{\alpha^2 M_{E_-^{\one }}^2} {24\pi \cos^{4}\!\theta_{w} \, M_{B_\mu^\one}} \, 
{\cal I}_-\!\left( M_{B_\mu^\one}, M_{B_H^\one}, M_{E_-^\one}\right)  ~,
\label{BmutoBh}
\ee
where ${\cal I}_-$ is another phase space integral defined in Eq.~(\ref{fpm}).
The width into left-handed leptons, 
\be
\Gamma\left(B^{\one }_\mu \to B^{\one }_{H} e_L^+e_L^-\right) 
= \frac{\alpha^2 M_{L_+^{\one }}^2} {384\pi \cos^{4}\!\theta_{w} \, M_{B_\mu^\one}} \, 
{\cal I}_-\!\left( M_{B_\mu^\one}, M_{B_H^\one}, M_{L_+^\one}\right)  ~,
\ee
is suppressed due to the smaller 
hypercharge and larger mass of the (1,0) fermion, which is $L_+^\one$ in this case.
For the same reasons, the $B^{\one }_\mu$ decay into a $B^{\one }_{H}$
and $q\overline{q}$ pairs has a small decay width.
$B^{\one }_\mu$ decays to $W^{\one }_H$ plus fermion pairs
are highly suppressed due to 
the dependence on the 7th power of the small difference between initial and 
final (1,0) masses [see Eqs.~(\ref{AmutoAH}) and (\ref{fpe}) in Appendix C].

Besides these tree-level 3-body decays,  $B^{\one }_\mu$ also has 2-body decays
via the dimension-5 operator shown in Eq.~(\ref{operator}), which is 
induced at one loop (see Appendix B). The decay width is given by
\be
\Gamma\left(B^{\one }_\mu \to B^{\one }_{H} \gamma \right) 
= \frac{\alpha^3}{96 \pi^2 \cos^{4}\!\theta_{w} } \frac{1}{M_{B_\mu^\one}}
\left(1- \frac{M_{B_H^\one}^2}{ M_{B_\mu^\one}^2} \right) 
\left( \sum_F  \sigma_F \, \Big ( \frac{Y_F}{2} \Big )^2 \, Q_F \, {\cal E}_F  \right)^2 ~,
\label{oneloopdecay}
\ee
where the sum over $F$ includes all quarks and leptons,
$\sigma_F$ is +1 for $SU(2)_W$ doublets and $-1$ for  $SU(2)_W$ singlets, 
$Q_F$ is the electric charge,
$Y_F$ is the hypercharge normalized 
to be twice the electric charge for $SU(2)_W$ singlets,
and  ${\cal E}_F$ is given in Eq.~(\ref{ef}) and depends only on 
the masses of $B_H^\one$, $B_\nu^\one$, and of the (1,0) and (1,1) fermions.
Using the values for the standard model gauge couplings given at the end of section 2.2,
{\it i.e.}, $\alpha = 1/127$ and $\sin^{2}\!\theta_{w} = 0.235$, 
we find the following branching fractions for  $B^{\one }_\mu$:
\bear
{\rm Br} \left(B^{\one }_\mu \rightarrow  B^{\one }_{H} \gamma \right) \equiv b_{B\gamma} 
\approx 34.0\% ~,
\nonumber \\ [ 2mm]
{\rm Br} \left(B^{\one }_\mu \rightarrow  B^{\one }_{H} e^+e^- \right) \equiv b_{Be} 
\approx 21.3\% ~.
\label{bmu-br} 
\eear
The branching fractions into $e^+e^-B^{\one }_{H}$, $\mu^+\mu^-B^{\one }_{H}$
and $\tau^+\tau^-B^{\one }_{H}$ are equal. 
The fact that the tree-level 3-body decay and the one-loop 2-body decay have
comparable  branching fractions in the case of $B^{\one }_\mu$
is an accidental consequence of the mass spectrum given in Table 1. 
The $B^{\one }_\mu$ 
decays into $B^{\one }_{H}$ plus neutrinos or quarks have small
branching fractions (1.4\% and 0.6\%, respectively) which may be 
safely ignored in what follows.

The (1,0) leptons can decay into (1,0) modes of the 
electroweak gauge bosons or spinless adjoints, and a standard model lepton.  
The decay widths of the $SU(2)_W$-doublet (1,0) leptons,
$L_+^{\one } \equiv (N_+^{\one }, E_+^{\one })$, to neutral (1,0) particles are given 
at tree level by: 
\bear
\Gamma\left(L_+^{\one }\!\to W_H^{\one 3} l_{L}\right) & \! =\! &
\frac{\alpha }{32\sin^{2}\!\theta_{w}}M_{L^{\one }}
\left(1-\frac{M_{W_H^{\one }}^2}{M_{L^{\one }}^2}\right)^{\! 2} ~,
\nonumber \\ [0.5em] 
\hspace{.1mm}
\Gamma\left(L_+^{\one }\!\to B^{\one }_{\mu} l_L \right) & \! =\! &
\frac{\alpha }{16 \cos^2\!\theta_{w}}  M_{L^{\one }}
\left(1 - \frac{M_{B_\mu^{\one }}^2}{M_{L^{\one }}^2} \right)^{\! 2} 
\left(1 + \frac{M_{L^{\one }}^2}{2M_{B_\mu^{\one }}^2}\right)~,
\nonumber \\ [0.5em] 
\hspace{.1mm} 
\Gamma\left(L_+^{\one }\!\to B_H^{\one } l_L \right)  & \! =\! &
\frac{\alpha }{32\cos^{2}\!\theta_{w}} M_{L^{\one }}
\left(1-\frac{M_{B_H^{\one }}^2}{M_{L^{\one }}^2}\right)^{\! 2} \, ,
\label{Ldecays}
\eear
where $l_L$ is the corresponding standard model weak doublet lepton.
The decays to charged (1,0) particles, $E_+^{\one }\!\to W_H^{\one -} \nu_L$ 
and $N_+^{\one }\!\to W_H^{\one -} e_L^+$, have a width twice as
large as the $L_+^{\one }\!\to W_H^{\one 3} l_{L}$ decay width.
The $L_+^{\one }$ branching fractions are given by:
\bear
&& 
{\rm Br} \left[ (N_+^{\one }, E_+^{\one }) \to B_H^{\one } (\nu_L,e_L) \right] 
\equiv b_{l1} \approx 20.1\%  ~.
\nonumber \\ [0.5em] 
&& 
\frac{1}{2}\, {\rm Br} \left[ (N_+^{\one }, E_+^{\one }) \to W_H^{\one +} (e_L,\nu_L) \right] 
=
{\rm Br} \left[ (N_+^{\one }, E_+^{\one }) \to W_H^{\one 3} (\nu_L,e_L) \right] 
\equiv b_{l2} \approx 23.5\% ~,
\nonumber \\ [0.5em] 
&& {\rm Br} \left[ (N_+^{\one }, E_+^{\one }) \to B_\mu^{\one }(\nu_L,e_L) \right] 
 \equiv b_{l3} \approx 9.3\% ~.
\label{onestep}
\eear

%
\TABLE[t]{
\centering
\renewcommand{\arraystretch}{1.7}
\begin{tabular}{|c|c|c|c|c|c|c|}
\cline{1-3}\cline{5-7}
\hspace{-0.1em} Final-state \hspace{-0.1em} &  
\multicolumn{2}{c|}{$W_\mu^{\one 3} \!\to ... \to B_H^{\one} $} & \hspace*{0.17em} & 
\hspace{-0.5em} Final-state \hspace{-0.7em} &  
\multicolumn{1}{r}
{$W_\mu^{\one +} \!\to ... \to B_H^{\one } $} & \rule{0mm}{5mm}\rule{0mm}{-22mm} 
\\ \cline{2-3}\cline{6-7} 
$e,\mu,\gamma$  & Branching fractions & \%  & & $e,\mu,\gamma$ & Branching fractions & \% 
\\  \cline{1-3}\cline{5-7}
$X$   & $\frac{\textstyle 2}{\textstyle 3}( b_{l1} +  b_{l2} + b_{l3}b_{Be} )$ & 30.4  & &
$X$   & $\frac{\textstyle 1}{\textstyle 3}(b_{l1} + 2 b_{l2} + b_{l3}b_{Be} )$ &  23.1
\\ \cline{1-3}\cline{5-7}
$(e^+ + e^-)X$        & $\frac{\textstyle 4}{\textstyle 9} b_{l2}$ & 10.5   & &
$e^+ \, X$           & $\frac{\textstyle 1}{\textstyle 3}(b_{l1} + 2 b_{l2} + b_{l3}b_{Be})$ & 23.1
\\ \cline{1-3}\cline{5-7}
$\!(e^+\!\mu^-\!\! + e^-\!\mu^+) X\!\!$ & $\frac{\textstyle 4}{\textstyle 9} b_{l2}$ &  10.5   & &
$e^+ e^- \, X $ & $\frac{\textstyle 1}{\textstyle 6} (b_{l2} + 2 b_{l3}b_{Be})$ &  \ 4.6
\\  \cline{1-3}\cline{5-7}
$e^+e^- \, X$              & 
$\frac{\textstyle b_{l1}}{\textstyle 6}+ \frac{\textstyle 4}{\textstyle 9} b_{l2} 
+ \frac{\textstyle 5}{\textstyle 6}b_{l3}b_{Be}\!$    & 15.5  & &
$\!e^+e^- e^+ X\!$ & $\frac{\textstyle 1}{\textstyle 6} (b_{l2} + 2 b_{l3}b_{Be})$ & \ 4.6
\\  \cline{1-3}\cline{5-7} 
$e^+e^- e^+e^-$           & $\frac{\textstyle 1}{\textstyle 36}(b_{l2} + 6 b_{l3} b_{Be} )$ & \ 1.0 & &
$\!e^+e^-\mu^+ X\!$ & $\frac{\textstyle 1}{\textstyle 6} (b_{l2} + 2 b_{l3}b_{Be} )$ & \ 4.6
\\  \cline{1-3}\cline{5-7}
$\!e^+e^- \mu^+\mu^-$       & $\frac{\textstyle 1}{\textstyle 18}(b_{l2} + 6 b_{l3}b_{Be})$ & \ 2.0 & &
$\gamma \, X$ & $\frac{\textstyle 1}{\textstyle 3} b_{l3}b_{B\gamma } $  & \ 1.1
\\  \cline{1-3}\cline{5-7}
$\gamma \, X$ & $\frac{\textstyle 2}{\textstyle 3} b_{l3}b_{B\gamma } $  & \ 2.1 & &
$\gamma \; e^+ \, X$ & 
$\frac{\textstyle 1}{\textstyle 3} b_{l3}b_{B\gamma } $  & \ 1.1
\\  \cline{1-3}\cline{5-7}
$\gamma \; e^+e^- \, X$ & $\frac{\textstyle 1}{\textstyle 6} b_{l3} b_{B\gamma } $ & \ 0.5 & 
\multicolumn{4}{c}{} 
\\  \cline{1-3} 
\end{tabular}
\medskip
\caption{Branching fractions for the complete cascade decays of $W_\mu^{\one 3}$ and 
 $W_\mu^{\one +}$. $X$ stands for a number of neutrinos or taus. 
The branching fractions involving more muons than electrons (not shown) are 
equal to the analogous ones involving more electrons than muons.
The branching fractions of $W_\mu^{\one -}$ are the same as for $W_\mu^{\one +}$
except for flipping the electric charges of the final state leptons. 
The branching fractions for `one-step' decays,  $b_{l1}$, $b_{l2}$, $b_{l3}$
and  $b_{Be}$, $b_{B\gamma }$, are defined in Eqs.~(\ref{onestep}) and (\ref{bmu-br}).\label{tab:totalBRW}}
}
%

As opposed to the three spinless adjoints and $B_\mu^{\one }$ 
which at tree level have only 3-body decays, 
the $W_\mu^{\one }$ particles are heavier than the (1,0) leptons 
and therefore decay with a branching fraction of almost 100\% into 
one (1,0) lepton doublet and the corresponding standard model lepton 
doublet.
Putting together the branching fractions for various decays of the electroweak 
(1,0) bosons, we find the  branching fractions for the  
complete cascade decays of $W_\mu^{\one 3}$ shown in Table \ref{tab:totalBRW}.

\subsection{Colored (1,0) particles}

At tree level, the (1,0) spinless adjoint of $SU(3)_c$ has only 3-body decays
into a quark-antiquark pair and one of the electroweak (1,0) bosons.
The decay widths are derived in Appendix C, and take the following form: 
\be
\Gamma\left(G^{\one }_H \to B^{\one }_{H} u_R\overline{u}_R\right) 
 =  \frac{y_{u_R}^2\alpha\alpha_s}{64\pi \cos^2\!\theta_w}\,
M_{G_H^{\one }} \, 
{\cal I}_+\!\left( M_{G_H^\one}, M_{B_H^\one}, M_{U_-^\one}\right)  ~,
\ee
\be
\Gamma\left(G^{\one }_H \to B^{\one }_\mu u_R\overline{u}_R\right) 
\approx 
\frac{y_{u_R}^2\alpha\alpha_s}{140\pi \cos^2\!\theta_w}\,
M_{G_H^{\one }} \, \frac{M_{U_-^{\one }}^2}{M_{B_\mu^{\one }}^2}
\frac{ \left( M_{G_H^{\one }} - M_{B_\mu^{\one }} \right)^7}
{( M_{U_-^{\one }}^2 - M^2_{G_H^{\one }} )^4} ~,
\ee
for hypercharge (1,0) bosons in the final state, and
\bear \hspace*{-3em}
\Gamma\left(G^{\one }_H \to W^{\one 3}_H u_L\overline{u}_L\right) 
& \approx & \frac{\alpha\alpha_s}{420\pi \sin^2\!\theta_w}\,
M_{G_H^{\one }}^2
\frac{ \left( M_{G_H^{\one }} - M_{W_H^{\one }} \right)^7}
{( M_{Q_+^{\one }}^2 - M^2_{G_H^{\one }} )^4} ~,
\nonumber \\ [0.6em] \hspace*{-3em}
\Gamma\left(G^{\one }_H \to W^{\one +}_H d_L \overline{u}_L  \right) 
& = & \Gamma\left(G^{\one }_H \to W^{\one -}_H u_L \overline{d}_L  \right) 
= 2 \, \Gamma\left(G^{\one }_H \to W^{\one 3}_H u_L \overline{u}_L \right) ~,
\eear
for $SU(2)_W$ (1,0) bosons. 
Note that we have expanded the decay widths to leading order in the mass difference 
of $G_H^{\one }$ and the electroweak (1,0) boson [see Eq.~(\ref{fpe})] 
in the case of $G_H\to B_\mu$ and $G_H\to W_H$ transitions, but not for 
$G_H\to B_H$ where the mass difference is larger and the expansion does not provide 
a good approximation.

$G^{\one }_H$ has also a two-body decay into $B^{\one }_\mu$ and a gluon,
via a dimension-5 operator shown in Eq.~(\ref{operator}), which is induced at one loop. 
However, the width for this decay is highly suppressed because 
$G^{\one }_H$ and  $B^{\one }_\mu$ are almost degenerate.

After summing over all quark flavors, we find that
the dominant decay mode of $G^{\one }_H$ is into $B^{\one }_H q\overline{q}$,
with a total branching fraction of $b_{g1} \approx 96.5\%$.
The sum over all branching fractions of $G^{\one }_H$ into $W^{\one +}_H$ or 
$W^{\one -}_H$ plus a quark-antiquark pair is $b_{g2}^\prime \approx 2.3\%$.
The branching fraction for $G^{\one }_H \to W^{\one 3}_H q\overline{q}$ 
is $b_{g2} \approx 1.2\%$, while the decay into $B_\mu^{\one }$ is highly suppressed due to the 
very small mass difference involved in that case.
The branching fractions quoted here correspond to $1/R = 500$ GeV. For different
values of $1/R$, the branching fractions of $G^{\one }_H$ change slightly due to
the dependence of $M_{T_\pm^\one}R$ on $1/R$ shown in Table~\ref{tab:mass}. 
For the coupling constants we use $\alpha_s = 0.107$, 
$\alpha = 1/127$ and $\sin^2\theta_w = 0.235$, which are the standard model
values at 500 GeV.

The (1,0) quarks can decay into both vector and spinless
modes. 
The largest decay width is into a $G_H^{\one }$
and a standard model quark:
\be
\Gamma\left(Q^{\one }\!\to G_H^{\one } q\right) = \frac{\alpha_s}{6}\,
M_{Q^{\one }} 
\left(1 - \frac{M_{G_H^{\one }}^2}{M_{Q^{\one }}^2} \right)^{\! 2} ~.
\label{Q-GHq}
\ee
The $SU(2)_W$-doublet (1,0) quarks can also decay into a standard-model quark, 
and an $SU(2)_W$ gauge boson or spinless adjoint.
Ignoring the standard-model quark mass, the decay width for the latter is
\be
\Gamma\left(Q_+^{\one }\!\to W_H^{\one 3} q_{L}\right) =
\frac{\alpha }{32\sin^{2}\!\theta_{w}}M_{Q_+^{\one }}
\left(1-\frac{M_{W_H^{\one }}^2}{M_{Q^{\one }}^2}\right)^{\! 2} ~,
\label{Q-WHq}
\ee
and is twice as large in the case of  $W_H^{\one \pm}$.
The decays of (1,0) quarks into an $SU(2)_W$ (1,0) vector boson
and a standard model quark have a width
\be
\Gamma\left(Q^{\one }_+\!\to W^{\one  3}_{\mu} q_{L} \right) =
\left(\frac{M_{Q_+^{\one }}^2 - M_{W_\mu^{\one }}^2}
{M_{Q_+^{\one }}^2 - M_{W_H^{\one }}^2}\right)^{\! 2}
\left(2 + \frac{M_{Q_+^{\one }}^2}{M_{W_\mu^{\one }}^2}\right) 
\Gamma\left(Q_+^{\one }\!\to W_H^{\one 3} q_{L}\right) ~.
\label{Q-Wq}
\ee
The width is twice as large for $Q_+^{\one }\!\to W^{\one \pm}_{\mu} q_{L}$. 

All (1,0) quarks may also decay into (1,0) hypercharge bosons with widths
\bear
\hspace{-1.1cm} 
\Gamma\left(Q^{\one }\!\to B_H^{\one } q\right)  & \! =\! &
\frac{Y^{2}_{q}\alpha }{32\cos^{2}\!\theta_{w}} M_{Q^{\one }}
\left(1-\frac{M_{B_H^{\one }}^2}{M_{Q^{\one }}^2}\right)^{\! 2} ~,
\nonumber \\ [0.6em] 
\hspace{-1.1cm} 
\Gamma\left(Q^{\one }\!\to B^{\one }_{\mu} q \right) & \! =\! &
\left(\frac{M_{Q^{\one }}^2 - M_{B_\mu^{\one }}^2}
{M_{Q^{\one }}^2 - M_{B_H^{\one }}^2}\right)^{\! 2}
\left(2 + \frac{M_{Q^{\one }}^2}{M_{B_\mu^{\one }}^2}\right) 
\Gamma\left(Q^{\one }\!\to B_H^{\one } q\right) ~,
\label{Q-Bq}
\eear
where $Y_{q}$ is the hypercharge of the quark $q$, normalized to be 1/3 for 
$SU(2)_W$ doublets.
The branching fractions of the (1,0) quarks of the first and second generations
are shown in Table \ref{tab:BRquarks}.

\begin{figure}[t]
\centerline{
\psfig{file=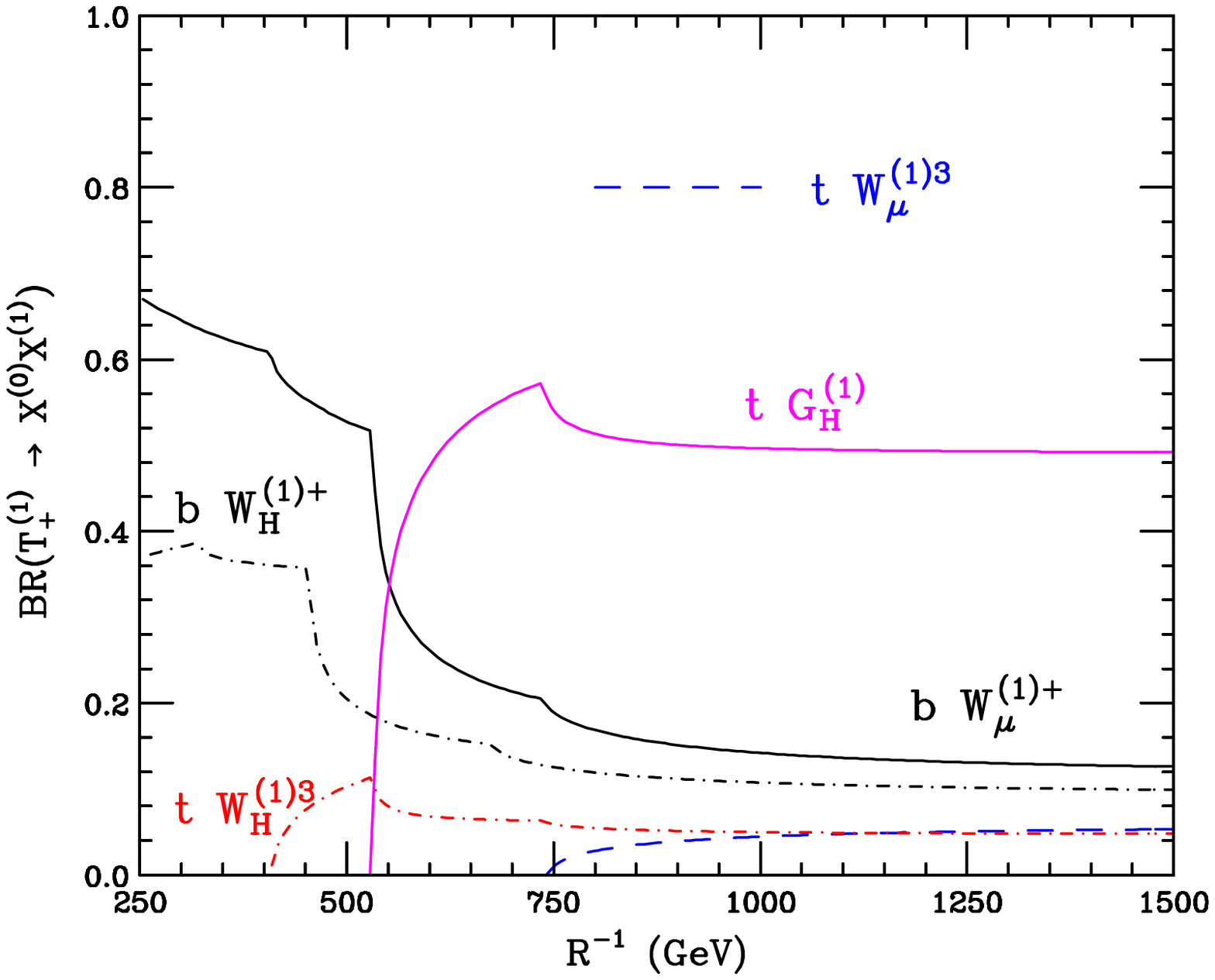,width=7.8cm,angle=0}
\psfig{file=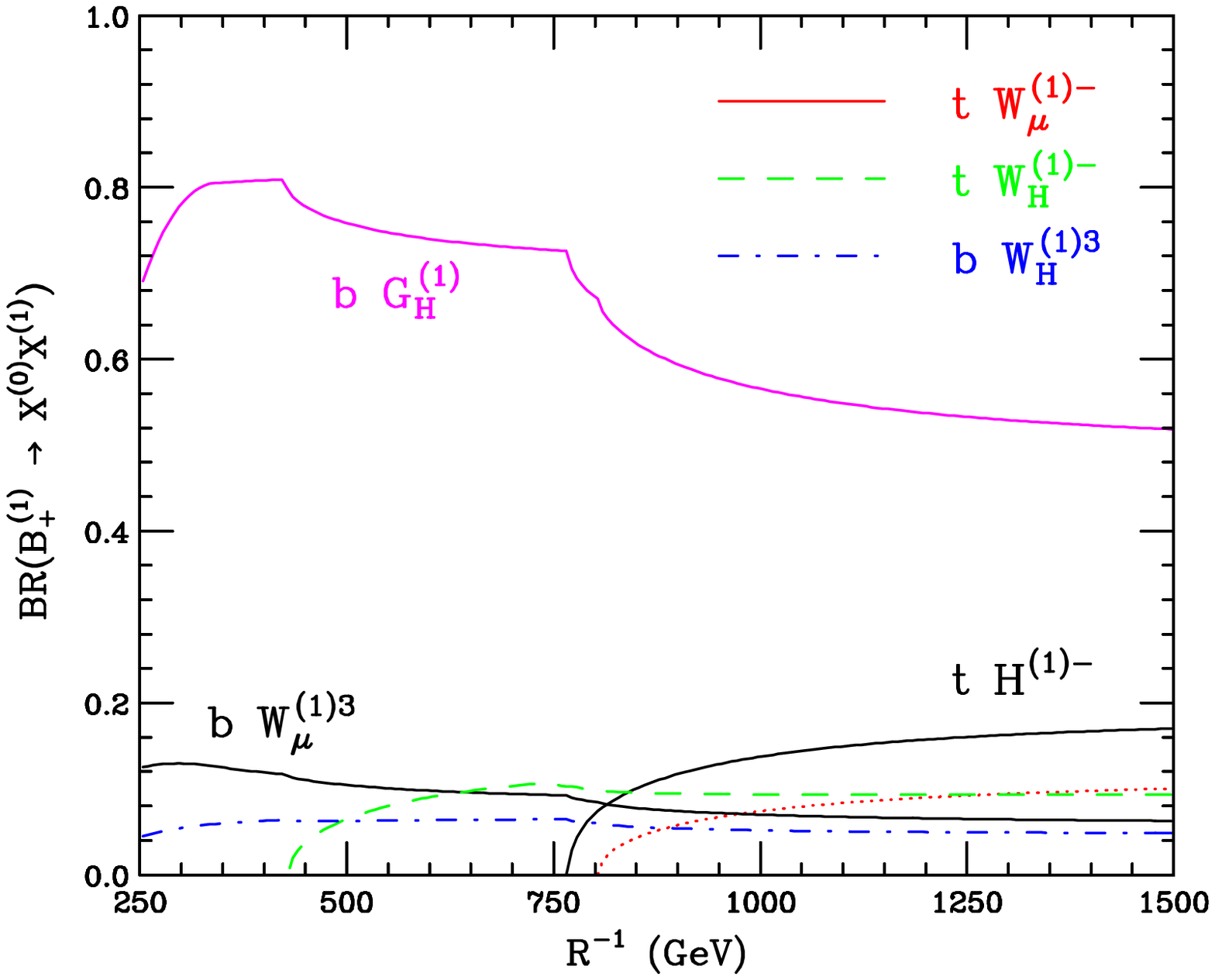,width=7.8cm,angle=0}
}
\vspace*{-1.4mm}
\caption{Branching fractions for the $SU(2)_W$-doublet (1,0) quarks of the third generation,
 assuming that the (1,0) Higgs particles have a mass $M_{H^\one} = 1.05/R$.
}
\label{fig:Brs}
\end{figure}

The $B_-^{\one }$ quark has the same branching fractions as $D_-^{\one }$,
while those of the $Q_+^{\one 3}=(T_+^{\one },B_+^{\one })$ 
 quarks are more sensitive to $1/R$, as shown in Figure~\ref{fig:Brs},
because of the large top quark mass.
Finally, the KK mode of the $SU(2)_W$-singlet top quark, $T_-^{\one }$,
has branching fractions highly sensitive to the mass of (1,0) Higgs
particles, with the decay into $b\, H^{\one +}$ dominating over $t\, G_H^\one$ 
if  $H^{\one +}$ is light.  Because of this fact, and also because of their small 
production cross section, third generation fermions do not result in many 
multi-lepton events.  Hence we will not give an expression for their branching fractions here.

\begin{table}[t]
\centering
\renewcommand{\arraystretch}{1.7}
\begin{tabular}{|c|c|c|c|c|c|}
\cline{1-2} \cline{4-6} 
\ $V^{\one }$ \ & Br$\left(U_+^{\one } \!\to q_L V^{\one }\!\right)\!$ & \hspace*{0.5em} & 
\ $V^{\one }$ \ & Br$\left(U_-^{\one } \!\to u_R V^{\one }\!\right)\!\!$ & 
Br$\left(D_-^{\one } \!\to d_R V^{\one }\!\right)\!\!\!\!$ 
\rule{0mm}{5mm}\rule{0mm}{-22mm} \\  \cline{1-2}\cline{4-6} \cline{1-2} \cline{4-6}
\cline{1-2}\cline{4-6} \cline{1-2} \cline{4-6}
$G_H^{\one }$ &  $b_{q3} \approx 63.2 \%$ & &
$G_H^{\one }$ &  $b_{u3} \approx 82.1\%$ &  $b_{d3} \approx 94.8\%$
\\ \cline{1-2} \cline{4-6} 
$\!\!W_\mu^{\one 3}$ ; $W_\mu^{\one +}\!\!$ & $b_{q2} \approx 6.4\%$ ; $2b_{q2}$ & &  
$B_\mu^{\one }$  & $b_{u2} \approx 11.5\%$ & \ \ $b_{d2} \approx 3.3\%$   
\\ \cline{1-2} \cline{4-6} 
$\!\!W_H^{\one 3}$ ; $W_H^{\one +}\!\!$ & $b_{q1} \approx 5.6\%$ ; $2b_{q1}$  &  &  
$B_H^{\one }$ & \ \ $b_{u1} \approx 6.4\%$ & \ \ $b_{d1} \approx 1.9\%$
\\ \cline{1-2} \cline{4-6} 
$B_\mu^{\one }$  & $b_{q0} \approx 0.55 \%$ & \multicolumn{3}{c}{} \\  \cline{1-2} 
\end{tabular}
\medskip
\caption{Branching fractions of first and second generation (1,0) quarks, in percentage. 
$D_+^\one$ have the same branching fractions as $U_+^\one$
except for a flip of the electric charge of the (1,0) bosons.
The $U_+^\one$ decays into a $B_H^{\one }$  and a quark is not shown 
because it is too small to be relevant.
}
\medskip
\label{tab:BRquarks}
\end{table} 

The (1,0) vector gluon decays into a standard model quark and a (1,0) 
quark. The width in the case of $SU(2)_W$-singlet down-type quarks
is given by
\be
\label{Gmu-width}
\Gamma\left(G_\mu^{\one }\! \rightarrow \sum_{i=1,2,3} D_{-_R}^{\one i} 
d_R^i\right) = 
\frac{\alpha_{s}}{2} M_{G_\mu^{\one }} 
\left(1-\frac{M_{D_-^{\one }}^2}{M_{G_\mu^{\one }}^2}\right)^{\!2} 
\left(1+\frac{M_{D_-^{\one }}^2}{2 M_{G_\mu^{\one }}^2}\right) ~.
\ee
The widths into all other (1,0) quarks except for the top have similar forms.
For $1/R \lae 1.3$ TeV the decays of the (1,0) vector gluon into 
$t_L T_{+_L}^{\one }$ or $t_R T_{-_R}^{\one }$ have a highly suppressed
phase space, and the branching fractions of 
$G_\mu^{\one }$ into a quark plus $Q_{+_L}^{\one i}$,
$U_{-_R}^{\one i}$, or $D_{-_R}^{\one i}$, summed over the index $i$ which labels 
the three generations, are given by 36.7\%, 24.6\% and 38.7\%, respectively.

For the purpose of analyzing the capability of the LHC to test
this model, we need to compute the branching fractions of the complete
cascade decays of the (1,0) quarks and gluons into the LKP and a
number of charged leptons or photons. 
It is useful to compute first the sums over branching fractions 
of the cascade decays that do not involve any $e^\pm$, $\mu^\pm$, or $\gamma$
for $G_H^{\one }$,
\be
b_{gX} = b_{g1} + \frac{2}{3}b_{g2} + \frac{b_{g2}^\prime}{3} ~,
\ee
and for $U_-^\one$, $D_-^\one$, $Q_+^\one$, respectively:
\bear
b_{uX} & = & b_{u1} + b_{u2}\, b_{Be} + b_{u3}\, b_{gX} ~,
\nonumber \\ [2.4mm]
b_{dX} & = & b_{d1}+ b_{d2}\, b_{Be} + b_{d3}\, b_{gX} ~,
\nonumber \\ [2.4mm]
b_{qX} & = &  b_{Be} b_{q0} +
\frac{4}{3} b_{q1} + \frac{2}{3}\left(2 b_{l1} + 3 b_{l2} + 2 b_{l3}b_{Be} \right) b_{q2} 
+ b_{q3}\, b_{gX} ~.
\eear
The right-hand sides of the above equations are sums over separate 
cascade decays, whose branching fractions are written as products
of `one-step' decays. For example, in the case of $b_{qX}$ the first term 
comes from the $Q_+^\one \to W_H^\one \to B_H^\one$ cascade, the second term comes from the 
sum over $Q_+^\one \to W_\mu^\one \to \cdots \to B_H^\one$ cascades, and the last term 
comes from the $Q_+^\one \to G_H^\one \to B_H^\one$ cascade.

The $Q_-^{\one}$ and $G_H^{\one }$ cascade
decays lead to at most two charged leptons, with small branching
fractions, as shown in Table~\ref{tab:BRGH}.
By contrast, $Q_+^{\one}$ 
have larger branching fractions for decays involving charged
leptons, and include up to four charged leptons (see Table~\ref{tab:BRGmu}).
However, the cascade decay with the largest branching fraction to a photon
is that of $U_-^{\one}$.

\begin{table}[t]
\centering
\renewcommand{\arraystretch}{1.7}
\begin{tabular}{|c||c|c|c|}
\hline
\hspace{-0.4em} Final-state $e,\mu,\gamma$ \hspace{-0.4em} &  
$G_H^{\one} \!\to ... \to B_H^{\one } $
&  $U_-^{\one} \!\to ... \to B_H^{\one } $ & $D_-^{\one} \!\to ... \to B_H^{\one } $
\rule{0mm}{5mm}\rule{0mm}{-22mm} 
\\  \hline\hline
$X$  & $ b_{gX} 
\approx 98.0\%$ 
& $b_{uX} \approx 89.4\%$ \ \ & $b_{dX} \approx 95.5\%$ \ \
\\ \hline
$e^+ \, (\mu^+) \,  X$ & $\frac{\textstyle 1}{\textstyle 6} b_{g2}^\prime \approx 0.38\%$ 
& $\frac{\textstyle 1}{\textstyle 6}b_{u3}b_{g2}^\prime \approx 0.31\%$ & 
$\frac{\textstyle 1}{\textstyle 6}b_{d3}b_{g2}^\prime \approx 0.36\%$
\\ \hline
$e^- \, (\mu^-) \, X$ & $\frac{\textstyle 1}{\textstyle 6} b_{g2}^\prime \approx 0.38\%$ 
& $\frac{\textstyle 1}{\textstyle 6}b_{u3}b_{g2}^\prime \approx 0.31\%$ & 
$\frac{\textstyle 1}{\textstyle 6}b_{d3}b_{g2}^\prime \approx 0.36\%$
\\ \hline
$e^+e^- \, (\mu^+\mu^-) \, X$  & $\frac{\textstyle 1}{\textstyle 6}b_{g2} \approx 0.21\%$ 
& $b_{u2}b_{Be}
+ \frac{\textstyle b_{u3}}{\textstyle 6}b_{g2} \approx 2.6\%$  & 
$b_{d2} b_{Be} + \frac{\textstyle b_{d3}}{\textstyle 6}b_{g2} \approx 0.90 \% $
\\ \hline
$\gamma \, X$ & $\approx 0$ & $b_{u2}b_{B\gamma} \approx 3.9\%$ & $b_{d2}b_{B\gamma} \approx 1.1\%$
\\ \hline
\end{tabular}
\medskip
\caption{Branching fractions for the complete cascade decays of $G_H^\one$,
$U_-^{\one}$ and $D_-^{\one}$, with 0,1 or 2 charged leptons in the final state.
$X$ stands for a number of standard model fermions other than $e^{\pm}$ 
and $\mu^\pm$.
The branching fractions for $\overline{U}_-^{\one}$ and $\overline{D}_-^{\one}$
are the same as for $U_-^{\one}$ and $D_-^{\one}$.
}
\label{tab:BRGH} 
\end{table} 

\newpage 

\vspace*{0.81cm}
\begin{table}[t!]
\centering
\renewcommand{\arraystretch}{1.7}
\begin{tabular}{|c|c|}
\hline
\hspace{-0.1em} Final-state $e,\mu ,\gamma $  & $U_+^{\one} \!\to ... \to B_H^{\one } $
\rule{0mm}{5mm}\rule{0mm}{-22mm} \\ \hline\hline
$X$ & $b_{qX} \approx 74.5\%$ 
\\ \hline
$e^+ \, (\mu^+) \, X$ & $\frac{\textstyle 2}{\textstyle 3}b_{q1} +
\frac{\textstyle 2}{\textstyle 9}\left(3 b_{l1} + 7 b_{l2} + 3 b_{l3} b_{Be} \right)b_{q2} 
+\frac{\textstyle 1}{\textstyle 6}b_{g2}^\prime b_{q3} \approx 7.3\% $  
\\ \hline
$e^- \, (\mu^-) \, X$ & \ \ $\frac{\textstyle 2}{\textstyle 9}b_{l2}b_{q2} 
+\frac{\textstyle 1}{\textstyle 6}b_{g2}^\prime b_{q3}\approx 0.58\% $ 
\\ \hline
$e^+e^- \, (\mu^+\mu^-) \, X$  
& $b_{Be} b_{q0} +
\frac{\textstyle b_{q1}}{\textstyle 6} +
\frac{\textstyle 1}{\textstyle 18}(3 b_{l1} + 14 b_{l2} + 27 b_{l3} b_{Be}  )b_{q2} 
+ \frac{\textstyle b_{g2}}{\textstyle 6} b_{q3} \approx 2.6\% $ 
\\ \hline
$e^+\mu^- \, (e^-\mu^+) \, X$ & $\frac{\textstyle 2}{\textstyle 9}b_{l2}b_{q2} \approx 0.33\% $ 
\\ \hline
$e^+e^+e^- \, (\mu^+\mu^+\mu^-) \, X$ 
& $\frac{\textstyle 1}{\textstyle 3} (b_{l2} + 2 b_{l3}  b_{Be} )b_{q2} \approx 0.58\%$ 
\\ \hline
$\mu^+e^+e^- \, (e^+\mu^+\mu^-) \, X$ 
& $\frac{\textstyle 1}{\textstyle 3} (b_{l2} + 2 b_{l3}  b_{Be} )b_{q2} \approx 0.58\%$ 
\\ \hline
$\!e^+e^-e^+e^- \, (\mu^+\mu^-\mu^+\mu^-)X\!\!$ 
& $\frac{\textstyle 1}{\textstyle 36}(b_{l2} + 6 b_{l3} b_{Be})b_{q2} \approx 0.063\%$ 
\\ \hline
$e^+e^-\mu^+\mu^- \, X$ 
& $\frac{\textstyle 1}{\textstyle 18}(b_{l2} + 6 b_{l3} b_{Be})b_{q2} \approx 0.13\%$ 
\\ \hline
$\gamma \, X$ & $b_{B\gamma } b_{q0} +
\frac{\textstyle 4}{\textstyle 3} b_{l3} b_{B\gamma } b_{q2} \approx 0.38\% $ 
\\ \hline
$\gamma \, e^+ \, (\gamma \mu^+ ) \, X$ & $\frac{\textstyle 2}{\textstyle 3} b_{l3} b_{B\gamma } b_{q2} \approx 0.13\%$  
\\ \hline
$\gamma \, e^+e^- \, (\gamma \mu^+\mu^- ) \, X$ & 
$\frac{\textstyle 1}{\textstyle 6} b_{l3} b_{B\gamma } b_{q2} \approx 0.033\% $ 
\\ \hline
\end{tabular}
\medskip
\caption{Branching fractions for the complete cascade decays of 
$U_+^{\one}$ with up to four charged leptons or photons in the final state.
$X$ stands for a number of standard model fermions other than $e^{\pm}$ 
and $\mu^\pm$. 
$\overline{D}_+^{\one}$ has the same branching fractions as $U_+^{\one}$, while 
the branching fractions of $D_+^{\one }$ and $\overline{U}_+^{\one}$ 
are given by flipping the lepton charges in the first column. 
The (1,0) top-quark doublet has braching fractions which are highly dependent 
on $1/R$, and are not shown here.
}
\label{tab:BRGmu}
\end{table}
%


\section{Signatures of (1,0) particles at hadron colliders}
\label{sec:production}

In this section we discuss the prospects for discovery of (1,0) particles at the LHC and the Tevatron. 
As shown in the previous section, a large number of leptons arises in the decays of 
$W_\mu^\one$ and 
other (1,0) bosons, while photons arise in the decay of the $B_\mu^\one$ vector boson.
We focus on computing the production cross sections of colored particles 
and the number of events with leptons and photons resulting from their decays.
We will also include direct production of $W_\mu^{\one}$ in our analysis although 
this turns out to have a rather small effect.

%
\subsection{Pair production of level-1 particles}

We discuss the production of (1,0) particles in order of importance 
for the lepton + photon signals under consideration.
This is more complicated than level-1 production in the case of 
one universal extra dimension \cite{Macesanu:2002db} 
because of the $G_H^\one$ spinless adjoint, which is not present in the 5D theory,
and appears in the final state as well as in $s$- and $t$- channel exchanges.

We begin with the $SU(2)_W$-doublet quark $Q_+^\one$, because 
a large fraction of its cascade decays gives rise to charged leptons (see 
Section~\ref{sec:decays}).
In addition, since it is lighter than the (1,0) vector gluon, 
and because of its high multiplicity, 
we expect $Q_+^\one$ production to be the dominant source of multi-lepton signals.
 We concentrate here on production mechanisms at the LHC, while
in section 4.3 we adapt this discussion to the case of $p\bar{p}$ collisions at the
Tevatron.

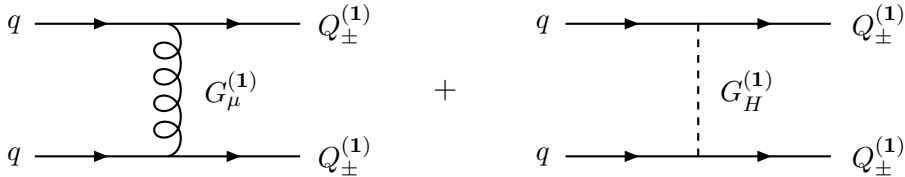
\begin{figure*}[b]
\unitlength=1.0 pt
\SetScale{1.0}
\SetWidth{0.8}      
\begin{center}
\begin{picture}(350,80)(40,20)
\ArrowLine( 60,75)(110,75)
\ArrowLine(60,25)(110,25)
\ArrowLine(110,25)(160,25)
\ArrowLine(110,75)(160,75)
\Gluon(110,75)(110,25){5}{4}
\Text(  220,50)[r]{{\large +}}
\ArrowLine(260,25)( 310,25)
\ArrowLine(310,25)( 360,25)
\ArrowLine(310,75)( 360,75)
\ArrowLine(260,75)( 310,75)
\DashLine(310,25)(310,75){3}
\Text(188,25)[r]{$ Q_\pm^\one$}
\Text(188,75)[r]{$ Q_\pm^\one$}
\Text(390,25)[r]{$ Q_\pm^\one$}
\Text(390,75)[r]{$ Q_\pm^\one$}
\Text(55,25)[r]{$ q$}
\Text(55,75)[r]{$ q$}
\Text(255,25)[r]{$q$}
\Text(255,75)[r]{$ q$}
\Text(145,50)[r]{$ G_\mu^\one$}
\Text(340,50)[r]{$ G_H^\one$}
\end{picture}
\end{center}
%
\caption{Diagrams for $Q_\pm^\one Q_\pm^\one$ production from quark-quark ($qq$)
initial state.}
\label{fig:qqQQ}
\end{figure*}

Given that there are more quarks than anti-quarks involved in proton-proton collisions, 
we first discuss quark-initiated pair production, 
$qq \to Q_\pm^\one {Q}_\pm^\one$, 
which is mediated by $G_\mu^\one$ and $G_H^\one$ exchange in the $t$ channel, 
as shown in Fig.~\ref{fig:qqQQ}. 
Two (1,0) quarks of different flavors ($Q_\pm^\one {Q'}_\pm^\one$), 
and an $SU(2)_W$ doublet-singlet pair ($Q_+^\one {Q'}_-^\one$) are produced in 
a similar way.

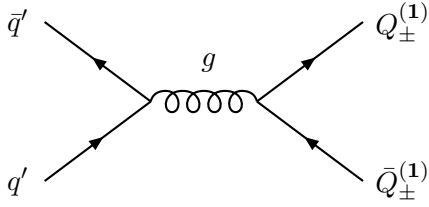
\begin{figure*}[t!]
\unitlength=1.0 pt
\SetScale{1.0}
\SetWidth{0.8}      
\begin{center}
\begin{picture}(160,80)(0,20)
\ArrowLine( 20,20)(  60,50)
\ArrowLine( 60,50)(  20,80)
\Gluon(60,50)(100,50){4}{4}
\ArrowLine( 140,20)(100,50)
\ArrowLine(100,50)( 140,80)
\Text(166,20)[r]{$\bar{Q}_\pm^\one$}
\Text(166,80)[r]{$ Q_\pm^\one$}
\Text(14,20)[r]{$ q'$}
\Text(14,80)[r]{$\bar{q}'$}
\Text(85,65)[r]{$ g$}
\end{picture}
\end{center}
%
\caption{$Q_\pm^\one \overline{Q}_\pm^\one$ production from $q'\bar{q}'$ initial state.}
\label{fig:qqQQ'}
\end{figure*}

For low $1/R$, the quark anti-quark and gluon initiated production mechanisms
are also important.
Production from a quark anti-quark pair, 
$q \bar{q}' \to Q_\pm^\one \bar{Q'}_\pm^\one$ 
and $q \bar{q}' \to Q_\pm^\one \bar{Q'}_\mp^\one$, is similar to the process shown 
in Fig.~\ref{fig:qqQQ} with a fermion line replaced by an anti-fermion line.
When quarks in the initial state have a different flavor than the (1,0) quarks 
in the final state, $q' \bar{q}' \to Q_\pm^\one \bar{Q}_\pm^\one$, 
a single tree-level diagram with a gluon exchange in the $s$ channel contributes,
as shown in Fig.~\ref{fig:qqQQ'}. 
The processes $q \bar{q} \to Q_\pm^\one \bar{Q}_\pm^\one$ 
(for which the initial and final states have same flavors) get contributions from 
the two diagrams in Fig.~\ref{fig:qqQQ} with one of the fermion lines replaced by 
an anti-fermion line, 
and also from the diagram of Fig.~\ref{fig:qqQQ'} with $q'$ replaced by $q$.

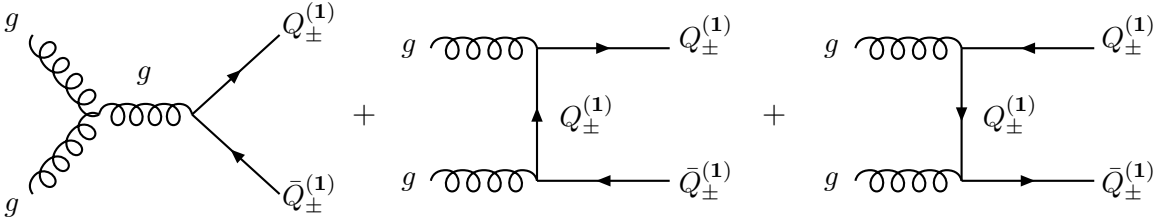
\begin{figure*}[h!]
\unitlength=1.0 pt
\SetScale{1.0}
\SetWidth{0.8}      
\begin{center}
\begin{picture}(440,80)(7,20)
\Gluon( 20,20)(  45,50){4}{4}
\Gluon( 45,50)(  20,80){4}{4}
\Gluon(45,50)(80,50){4}{4}
\ArrowLine( 113,20)(80,50)
\ArrowLine(80,50)( 113,80)
\Text(150,50)[r]{{\large +}}
\Gluon(170,75)(210,75){4}{4}
\Gluon(170,25)(210,25){4}{4}
\ArrowLine(210,25)(210,75)
\ArrowLine(210,75)(260,75)
\ArrowLine(260,25)(210,25)
\Text(305,50)[r]{{\large +}}
\Gluon(330,75)(370,75){4}{4}
\Gluon(330,25)(370,25){4}{4}
\ArrowLine(370,75)(370,25)
\ArrowLine(370,25)(420,25)
\ArrowLine(420,75)(370,75)
\Text(15,15)[r]{$g$}
\Text(15,85)[r]{$g$}
\Text(165,25)[r]{$ g$}
\Text(165,75)[r]{$ g$}
\Text(325,25)[r]{$g$}
\Text(325,75)[r]{$ g$}
\Text(65,65)[r]{$ g$}
\Text(135,20)[r]{$\bar{Q}_\pm^\one$}
\Text(135,85)[r]{$ {Q}_\pm^\one$}
\Text(285,25)[r]{$ \bar{Q}_\pm^\one$}
\Text(285,80)[r]{$ {Q}_\pm^\one$}
\Text(445,25)[r]{$\bar{Q}_\pm^\one$}
\Text(445,80)[r]{${Q}_\pm^\one$}
\Text(240,50)[r]{${Q}_\pm^\one$}
\Text(400,50)[r]{${Q}_\pm^\one$}
\end{picture}
\end{center}
%
\caption{Diagrams for $Q_\pm^\one \bar{Q}_\pm^\one$ production from gluon-gluon ($gg$)
initial state.}
\label{fig:ggQQ}
\end{figure*}

$Q_\pm^\one \bar{Q}_\pm^\one$ can also be produced from two gluons in 
the initial state, as shown in Fig.~\ref{fig:ggQQ}.
This production channel becomes increasingly important for smaller 
(1,0) quark mass (smaller $1/R$)
due to the larger gluon flux in the parton distribution.

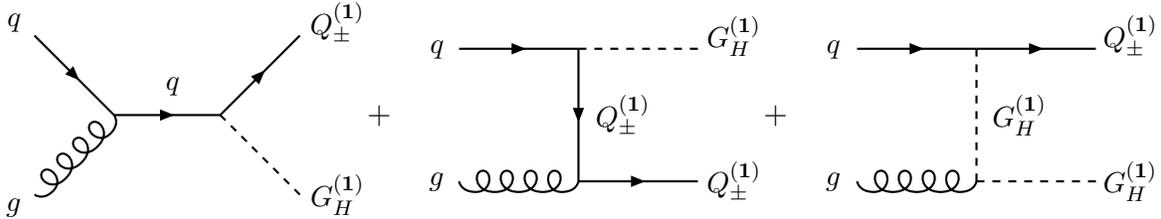
\begin{figure*}[b!]
\unitlength=1.0 pt
\SetScale{1.0}
\SetWidth{0.8}      
\begin{center}
\begin{picture}(450,80)(3,20)
\Gluon(50,50)( 20,20){4}{4}
\ArrowLine(  20,80)( 50,50)
\ArrowLine(50,50)(90,50)
\DashLine( 120,20)(90,50){3}
\ArrowLine(90,50)( 120,80)
\Text(155,50)[r]{{\large +}}
\ArrowLine(180,75)(225,75)
\ArrowLine(225,75)(225,25)
\Gluon(225,25)(180,25){4}{4}
\DashLine(270,75)(225,75){3}
\ArrowLine(225,25)(270,25)
\Text(305,50)[r]{{\large +}}
\ArrowLine(330,75)(375,75)
\Gluon(375,25)(330,25){4}{4}
\DashLine(375,75)(375,25){3}
\DashLine(375,25)(420,25){3}
\ArrowLine(375,75)(420,75)
\Text(15,15)[r]{$ g$}
\Text(15,85)[r]{$ q$}
\Text(175,25)[r]{$ g$}
\Text(175,75)[r]{$ q$}
\Text(325,25)[r]{$ g$}
\Text(325,75)[r]{$ q$}
\Text(75,61)[r]{$ q$}
\Text(145,20)[r]{$ G_H^\one$}
\Text(145,85)[r]{$ Q_\pm^\one$}
\Text(295,25)[r]{$ Q_\pm^\one$}
\Text(295,80)[r]{$ G_H^\one$}
\Text(445,25)[r]{$ G_H^\one$}
\Text(445,80)[r]{$ Q_\pm^\one$}
\Text(253,50)[r]{$ {Q}_\pm^\one$}
\Text(403,50)[r]{$ G_H^\one$}
\end{picture}
\end{center}
%
\caption{Diagrams for $G_H^\one Q_\pm^\one$ production from quark-gluon initial state.}
\label{fig:qgQGH}
\end{figure*}

Since the $SU(3)_c$ (1,0) bosons, $G_\mu^\one$ and $G_H^\one$, decay to fewer leptons 
than $Q_+^\one$, we will next consider their associated production with $Q_+^\one$.
The process $q g \to Q_\pm^\one G_H^\one$ is shown in Fig.~\ref{fig:qgQGH}. 
Diagrams with a (1,0) vector gluon in the final state can be 
obtained by replacing $G_H^\one$ by $G_\mu^\one$. 
Similar diagrams, but with  $G_H^\one$ replaced by $W_\mu^\one$ and an appropriate
flip between the up-type and down-type quarks, contribute to $q g \to Q_\pm^\one W_\mu^\one$
associated production.
%

$G_H^\one$ pair production is a
rather meager source of leptons or photons,
but for the sake of completeness we include here its diagrams:
quark initiated production $q\bar{q} \to G_H^\one G_H^\one$, and 
gluon initiated production $gg \to G_H^\one G_H^\one$ are shown in 
Figs.~\ref{fig:qqGHGH} and~\ref{fig:ggGHGH}, respectively. 
$G_\mu^\one$ pair production proceeds through the same diagrams
with all $G_H^\one$ lines replaced by $G_\mu^\one$ ones. 

$G_\mu^\one G_H^\one$ associated production, $q\bar{q} \to G_H^\one G_\mu^\one$, 
proceeds through four diagrams with $Q_+^\one$ and $Q_-^\one$ in the 
$t$ and $u$ channels, similar to the second diagram in Fig.~\ref{fig:qqGHGH}.
There is no contribution from the $s$ channel because the coupling 
$G_H^\one g^\mu G_\mu^\one$ does not exist at tree level due to gauge invariance. 

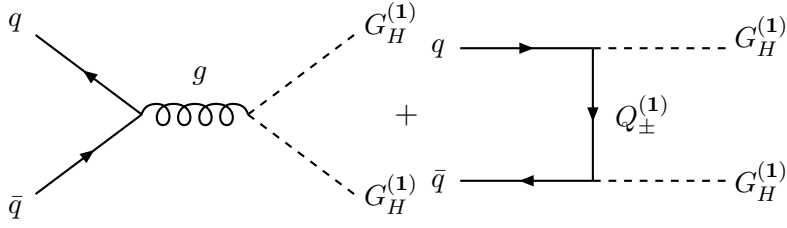
\begin{figure*}[t!]
\unitlength=1.0 pt
\SetScale{1.0}
\SetWidth{0.8}      
\begin{center}
\begin{picture}(300,80)(0,20)
\ArrowLine( 20,20)(  60,50)
\ArrowLine( 60,50)(  20,80)
\Gluon(60,50)(100,50){4}{4}
\DashLine( 140,20)(100,50){3}
\DashLine(100,50)( 140,80){3}
\Text(165,50)[r]{{\large +}}
\ArrowLine(180,75)(230,75)
\ArrowLine(230,25)(180,25)
\ArrowLine(230,75)(230,25)
\DashLine(280,75)(230,75){3}
\DashLine(280,25)(230,25){3}
\Text(15,15)[r]{$ \bar{q}$}
\Text(15,85)[r]{$ q$}
\Text(175,25)[r]{$ \bar{q}$}
\Text(175,75)[r]{$ q$}
\Text(85,65)[r]{$ g$}
\Text(165,20)[r]{$ G_H^\one$}
\Text(165,85)[r]{$ G_H^\one$}
\Text(305,25)[r]{$ G_H^\one$}
\Text(305,80)[r]{$ G_H^\one$}
\Text(260,50)[r]{$ {Q}_\pm^\one$}
\end{picture}
\end{center}
%
\caption{Diagrams for $G_H^\one G_H^\one$ production from $q\bar{q}$ 
( $u$-channel diagram is not shown).}
\label{fig:qqGHGH}
\end{figure*}
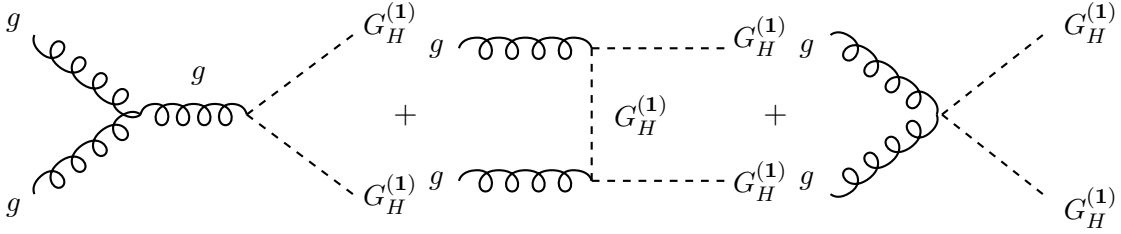
\begin{figure*}[t!]
\unitlength=1.0 pt
\SetScale{1.0}
\SetWidth{0.8}      
\begin{center}
\begin{picture}(450,80)(0,20)
\Gluon( 20,20)(  60,50){4}{4}
\Gluon( 60,50)(  20,80){4}{4}
\Gluon(60,50)(100,50){4}{4}
\DashLine( 140,20)(100,50){3}
\DashLine(100,50)( 140,80){3}
\Text(165,50)[r]{{\large +}}
\Gluon(180,75)(230,75){4}{4}
\Gluon(180,25)(230,25){4}{4}
\DashLine(230,25)(230,75){3}
\DashLine(230,75)(280,75){3}
\DashLine(280,25)(230,25){3}
\Text(305,50)[r]{{\large +}}
\Gluon(360,50)(320,20){4}{4}
\Gluon(320,80)(360,50){4}{4}
\DashLine(362,50)(400,20){3}
\DashLine(362,50)(400,80){3}
\Text(15,15)[r]{$ g$}
\Text(15,85)[r]{$ g$}
\Text(175,25)[r]{$ g$}
\Text(175,75)[r]{$ g$}
\Text(315,25)[r]{$ g$}
\Text(315,75)[r]{$ g$}
\Text(85,65)[r]{$ g$}
\Text(165,20)[r]{$ G_H^\one$}
\Text(165,85)[r]{$ G_H^\one$}
\Text(305,25)[r]{$ G_H^\one$}
\Text(305,80)[r]{$ G_H^\one$}
\Text(260,50)[r]{$ G_H^\one$}
\Text(430,15)[r]{$ G_H^\one$}
\Text(430,85)[r]{$ G_H^\one$}
\end{picture}
\end{center}
%
\caption{Diagrams for $G_H^\one G_H^\one$ production from $gg$
(a $u$-channel diagram is not shown).}
\label{fig:ggGHGH}
\end{figure*}

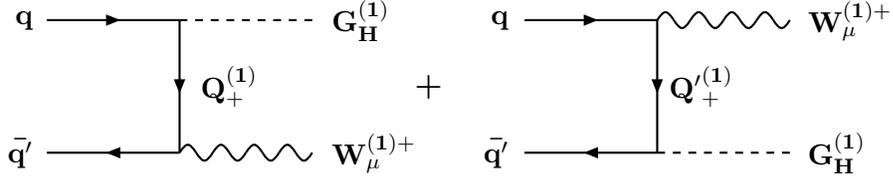
\begin{figure*}[h!]
\unitlength=1.0 pt
\SetScale{1.0}
\SetWidth{0.8}      
\begin{center}
\begin{picture}(350,80)(40,20)
\ArrowLine(80,75)(130,75)
\ArrowLine(130,25)(80,25)
\ArrowLine(130,75)(130,25)
\DashLine(180,75)(130,75){3}
\Photon(180,25)(130,25){3}{4}

\Text(230,50)[r]{{\bf\large +}}

\ArrowLine(260,75)(310,75)
\ArrowLine(310,25)(260,25)
\ArrowLine(310,75)(310,25)
\DashLine(360,25)(310,25){3}
\Photon(310,75)(360,75){3}{4}

\Text(75,25)[r]{$\bf \bar{q'}$}
\Text(75,75)[r]{$\bf q$}
\Text(255,25)[r]{$\bf \bar{q'}$}
\Text(255,75)[r]{$\bf q$}

\Text(220,25)[r]{$\bf W_\mu^{\one+}$}
\Text(210,75)[r]{$\bf G_{H}^\one$}
\Text(390,25)[r]{$\bf G_{H}^\one$}
\Text(400,75)[r]{$\bf W_\mu^{\one+}$}
\Text(160,50)[r]{$\bf {Q}_+^\one$}
\Text(340,50)[r]{$\bf {Q'}_+^\one$}
\end{picture}
\end{center}
%
\caption{Diagrams for $W_\mu^{\one+} G_H^\one$ production from $q\bar{q'}$.}
\label{fig:qqW+GH}
\end{figure*}

Finally we consider associated production of $G_{\mu}^\one$ or $G_H^\one$ with 
an $SU(2)_W$ vector boson, $W_\mu^{\one}$, 
as shown in Fig.~\ref{fig:qqW+GH} 
(with $G_H^\one$ in the final state replaced by $G_\mu^\one$ for 
$q \bar{q}' \rightarrow G_\mu^\one W_\mu^\one$).
For $W_\mu^{\one 3}$ in the final state, the initial state and the (1,0) 
quarks are all of the same type.
Associated production with hypercharge bosons, $B_\mu^\one$ $B_H^\one$, as 
well as with the $SU(2)_W$ spinless adjoints $W_H^\one$ are very small and 
will be neglected; 
we will also ignore production of (1,0) Higgs particles since their 
phenomenology is highly model-dependent.
%
%

Given that there are many diagrams that need to be taken into account, 
we have implemented the 6DSM detailed in section 2 
in {\tt CalcHEP}~\cite{Pukhov:1999gg,Pukhov:2004ca}, 
a tree-level Feynman diagram calculator (for a description of our 
{\tt CalcHEP} files, see \cite{web}).
Consequently it is rather straightforward to compute production cross sections for 
(1,0) particles at various colliders.
As a cross-check we have compared the {\tt CalcHEP} output for all 2- 
and 3-body decay widths with the corresponding analytic expressions 
in Section~\ref{sec:decays}. 
We also checked cross sections for selected production channels 
using {\tt MadGraph/MadEvent}~\cite{Stelzer:1994ta,Maltoni:2002qb}.

\begin{figure}[t]
\centerline{
\psfig{file=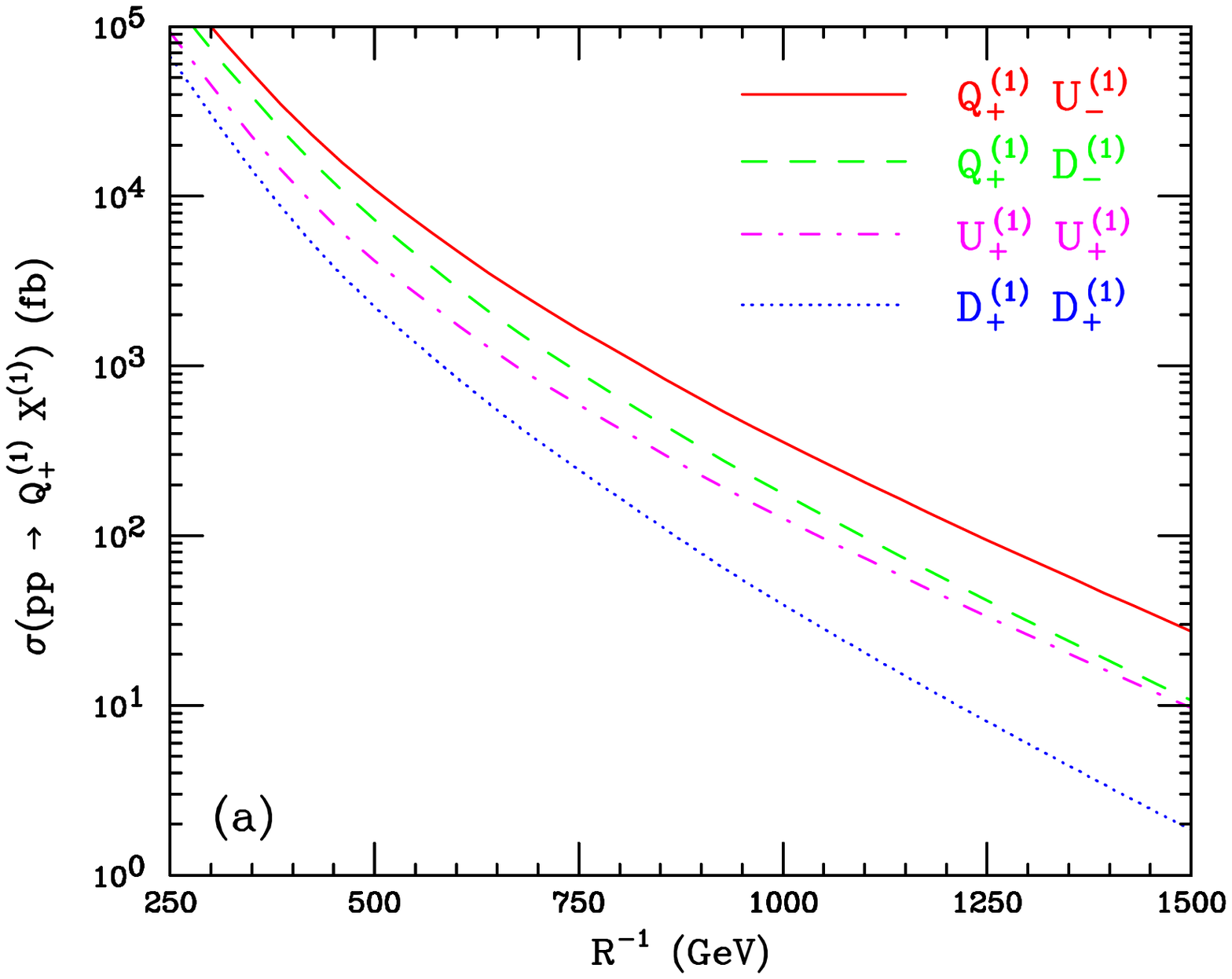,width=8cm,angle=0}
\psfig{file=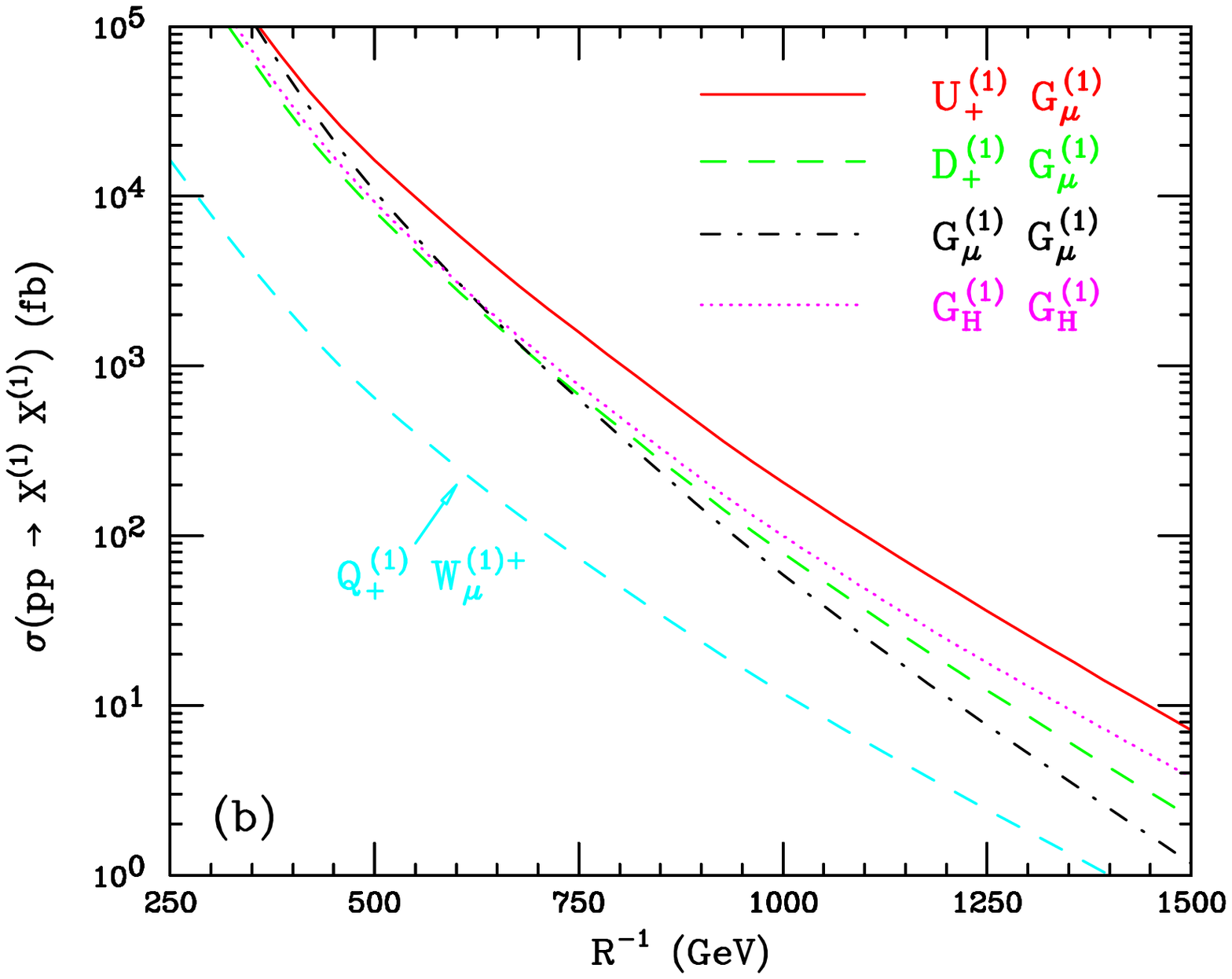,width=8cm,angle=0}  }
\vspace*{0.1mm}
\caption{Tree-level production cross sections of (1,0) particles at the LHC:  (a) quark pairs, and 
(b) final states involving bosons. The cross sections have been summed over the first two generations of 
KK quarks and antiquarks. The weak-doublet $Q^{\one}_+$ includes 
both up- and down-type (1,0) quarks. The cross section for $U^\one_+ D^\one_+$ production (not shown) 
turns out to be nearly equal to that for $U^\one_+ U^\one_+$. Cross sections 
for the weak-singlet quarks (6D chirality $-$) are almost the same as those 
for weak-doublet quarks (6D chirality $+$) and are not plotted.}
\label{fig:xsec_su3}
\end{figure}

The cross sections at the LHC ($\sqrt{s}=14$ TeV) are graphed as a function of 
$1/R$ in Fig.~\ref{fig:xsec_su3}, 
and have been summed over various channels.
We assume five partonic quark flavors in the proton along with the gluon, and 
ignore electroweak production of colored particles.
We use the CTEQ6L parton distributions~\cite{Pumplin:2002vw}, and 
choose the scale of the strong coupling 
constant $\alpha_s$ to be equal to the parton-level center of mass energy.

$Q_+^\one Q_+^\one$ production, which is responsible for most 
of the multi-lepton events (as shown later in Section~\ref{sec:lhc2}), 
is dominated 
by (1,0) quarks of the first 2 generations (88\% at $1/R=500$ GeV, increasing to 98\% at $1/R=1$ TeV). 
The gluon-gluon initial state contributes 
only $\sim$10\% (3\%) of the total $Q_+^\one Q_+^\one$ cross section at $1/R=500$ GeV
($1$ TeV), since firstly
the gluon flux in the proton at this 
mass scale is small, and secondly, there is a large number of subprocesses 
with $qq$ or $q\bar{q}$ initial states.
$G_H^\one$ production is different in that the dominant contribution to this process 
comes from the gluon initial state, with valence quarks making up 
the remainder.

The production cross sections of the $SU(2)_W$ doublet and singlet (1,0) 
quarks, $Q_+^{\one}$ or $Q_-^{\one}$, are almost equal, 
since they are produced in exactly the same way (see  Figs.~\ref{fig:qqQQ}-\ref{fig:qgQGH}).
The slightly higher mass of $Q_+^{\one}$ lowers its production cross section, but this is a small effect.
As expected from the structure of the parton distribution function, the $G_\mu^{\one}$ 
associated production cross sections drop off faster than others.

$Q_+^\one U_-^\one$ pair production, the main source of events containing 
both photons and leptons,
proceeds through $G_\mu^\one$ and $G_H^\one$ exchange in the $t$-channel,
as in Fig.~\ref{fig:qqQQ} with one of the $Q_+^\one$ quarks replaced by $U_-^\one$.
Due to the partonic structure, the production with first-generation   
quarks in the initial state are dominant, accounting for $\sim 50\%$ 
of all  $Q_+^\one U_-^\one$ pairs produced for $1/R=500$ GeV.

As mentioned earlier, $W_\mu^{\one}$ associated production, although small 
compared to that for colored (1,0) particles, is not necessarily negligible because
of its large branching fraction into leptons. We have included the 
cross section for the channel with the largest production rate, 
$W_\mu^{\one +} Q_+^{\one}$, in Fig.~\ref{fig:xsec_su3}.
The dominant contribution to this process is from production with first generation (1,0) quarks. 
$W_\mu^{\one -}$ associated production is even smaller, by an extra factor of $\sim$3, 
due to the partonic structure of the proton.

\subsection{Events with leptons and photons at the LHC}
\label{sec:lhc2}
%
Having determined the production rates of (1,0) particles, we now turn to a discussion 
of their experimental signatures at the LHC.
First we will consider the production of 
(1,0) particles which give $n\ell + m\gamma + \met$ with $n\geq n_{min}$ and $0 \leq m \leq 2$, 
where we do not count leptons from the decay of the standard model particles.

We calculate the inclusive cross sections for the channels $n\ell + m\gamma+\met$ with $n\geq 
n_{min}$ and $0 \leq m \leq 2$ 
in the following way. 
There are 11 (1,0) particles with different branching fractions to multiple leptons 
as discussed in Section~\ref{sec:decays}. We label these particles by $A_i^\one$, 
where $1 \leq i \leq 11$ is the particle type:
\begin{equation}
A_i^\one = \Big ( 
             W_\mu^{\one}, 
             G_\mu^{\one}, G_H^{\one}, T_+^{\one}, B_+^{\one}, T_-^{\one},
             U_-^{\one}, D_-^{\one}, 
	     Q_+^{\one} 
 \Big ) \, .
\end{equation}
Their branching fractions, Br$(i,a,a')$, where $a$ is the number of leptons 
($0 \leq a \leq 4$) and $a'$ is the number of photons ($0 \leq a' \leq 1$), are 
given in Section \ref{sec:decays}. $Q_+^{\one}$ and $U^\one_-$ include
only the first two generations of weak doublets and up-type singlets.  
One should keep in mind that the 3rd generation KK quarks and KK quarks of 
the first two generations have different branching fractions to leptons so 
they need to be tackled separately.  For simplicity we use the same symbol here
for quarks and antiquarks.
The cross section for $n\ell+m\gamma + \met$ events with $n\geq n_{min}$ and $0 \leq m \leq 2$ is 
\begin{equation}
\sigma (pp \to n\ell+m\gamma+\met~, n\geq n_{min} ) 
      = \sum_{i=1}^{11}\sum_{j\geq i}^{11} 
                 \sigma ( pp\to A_i^\one A_j^\one) B_{ij}  \, ,
\end{equation}
where $B_{ij}$ is a sum over products of branching fractions of the particles 
$A_i^\one$ and $ A_j^\one$
\begin{equation}
B_{ij} = \sum_{\underset{a+b\geq n_{min}}{a,b=0}}^{4} 
               \sum_{\underset{a'+b' = m}{a',b'=0}}^{1}
                  {\rm Br}(i,a,a'){\rm Br}(j, b,b') \, ,
\end{equation}
%
Note that the total numbers of photons ($m$) and leptons ($n$) from the decay of a pair
of (1,0) particles are constrained by 
$0 \leq n + 2m \leq 8$.  It is not possible to obtain $8\ell+2\gamma + \met$ for instance, 
since the hypercharge gauge boson $B_\mu^\one$ can decay into either a photon or a fermion pair, 
together with $B_H^\one$, so a photon is only produced at the expense of two leptons.
Most (1,0) particles have branching fractions that are independent of $1/R$.  However,
those for third generation quarks have variations due to 
threshold effects (see Fig.~\ref{fig:Brs}). 
We use values at large $1/R$, which 
slightly underestimates the total number of events as branching fractions are larger at small $1/R$.  Since the contribution from the third generation is small, 
our approximation gives rise to negligible error.
\begin{figure}[t]
\centerline{
\psfig{file=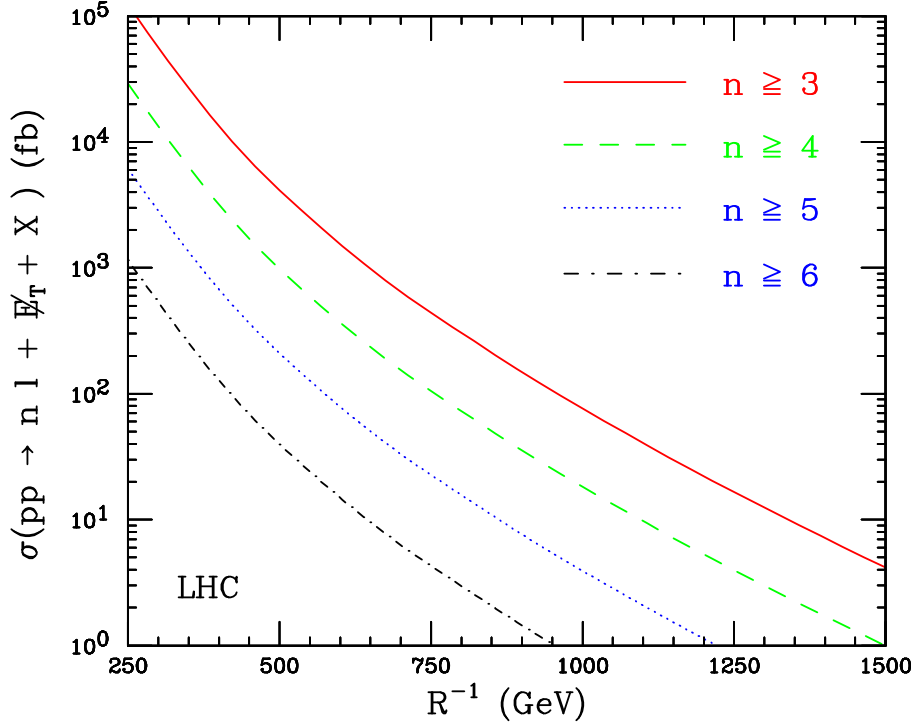,width=12cm,angle=0}
}
\vspace*{1mm}
\caption{Sum over cross sections for (1,0) particle pair production at 
the LHC times the branching fractions of 
the cascade decays that give rise to $n\geq 3,4,5$ or 6 charged leptons 
($\ell = e^\pm$ or $\mu^\pm$), 
 as a function of the compactification scale.} 
\label{fig:nevt}
\end{figure}
%
%

%
%
\begin{figure}
\centerline{
\psfig{file=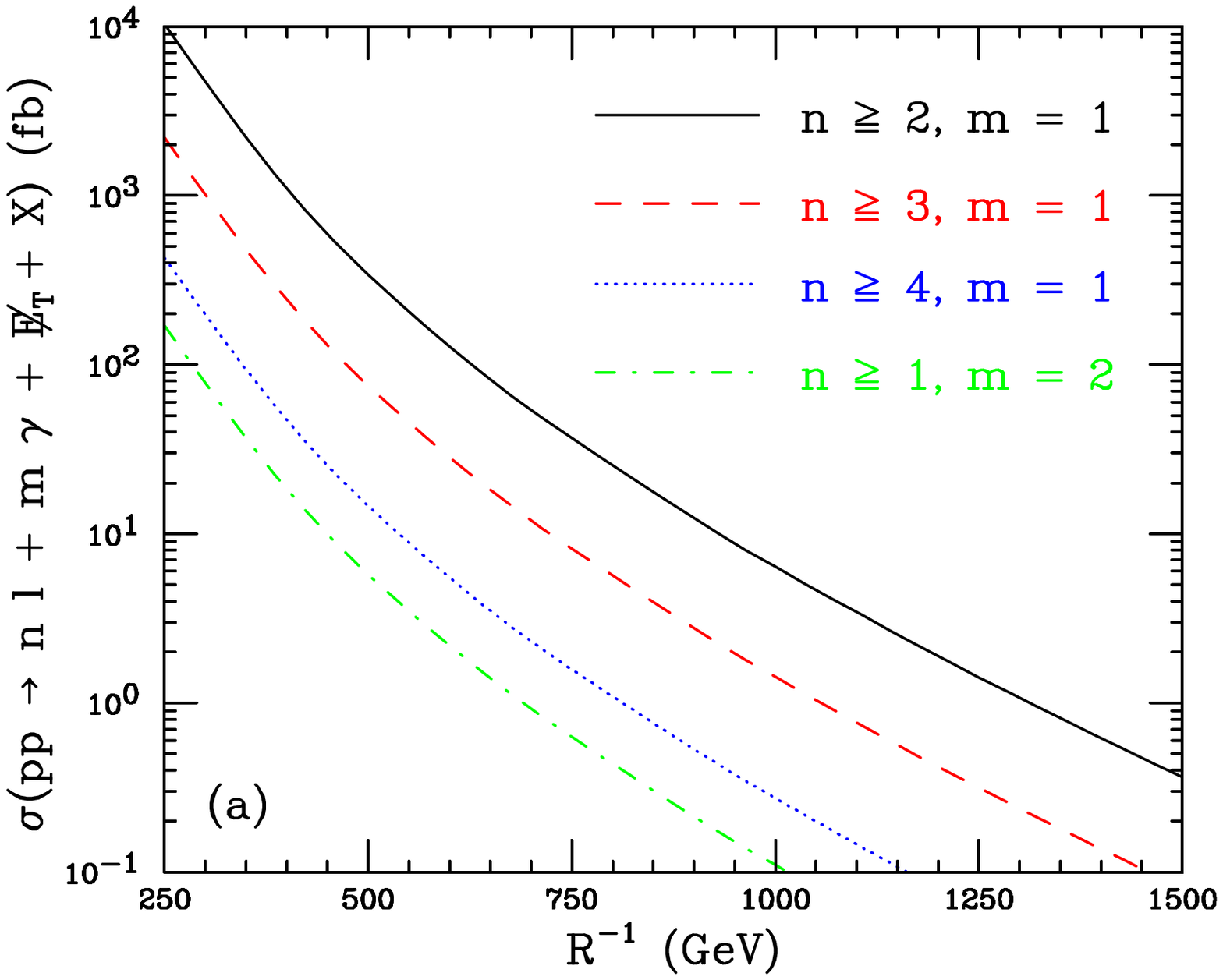,width=8cm,angle=0}
\psfig{file=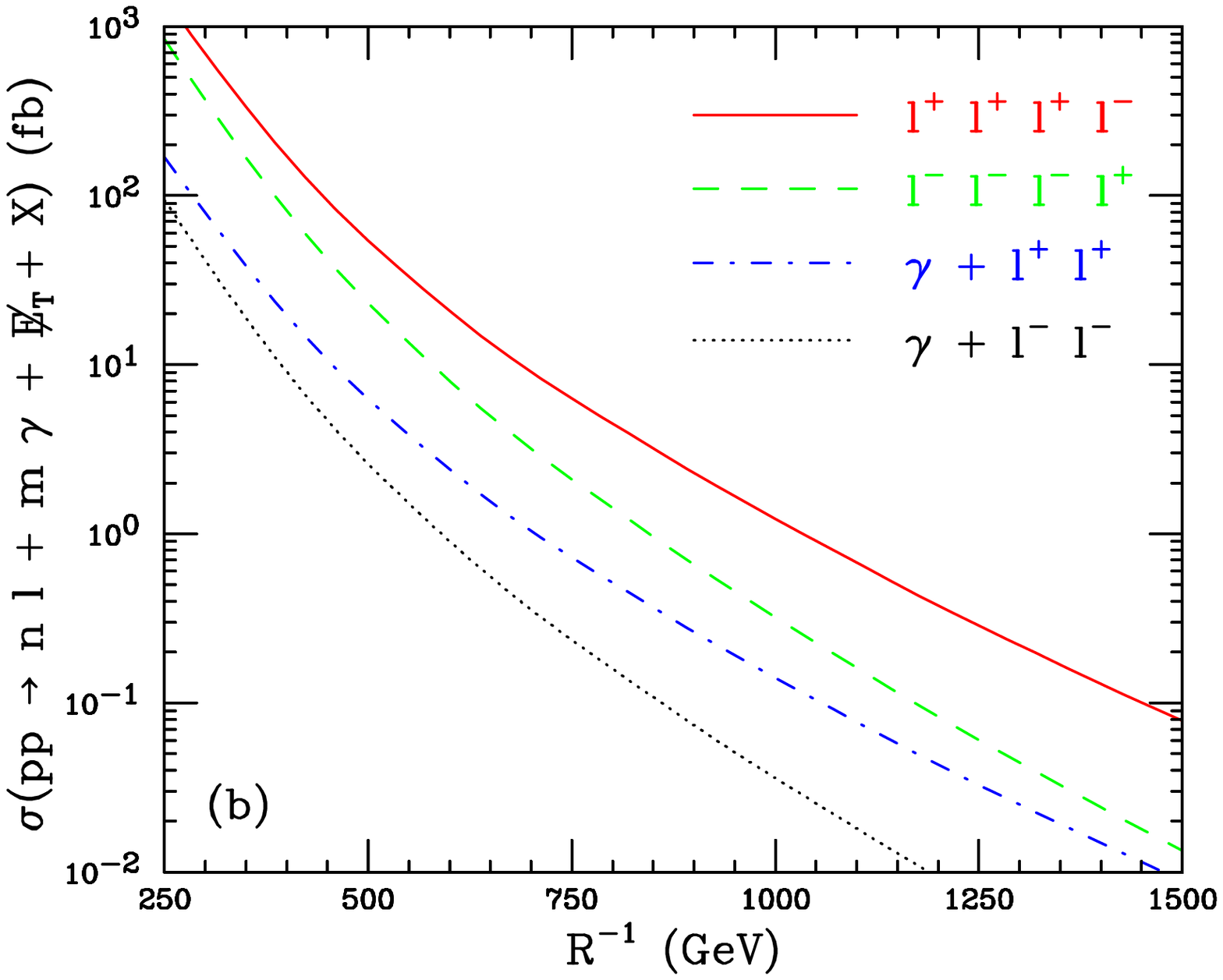,width=8cm,angle=0}
}
\vspace*{-4mm}
\caption{Cross sections for 
(a) $m\gamma+n\ell+\met$ events with $n\geq n_{min}$ for $m=1,2$ and $1 \leq n_{min} \leq 4$ 
and (b) Lepton + photon events with two or more same-sign leptons, at the LHC as a function of $1/R$.}
\label{fig:eeee}
\end{figure}
Cross sections for multi-lepton events at the LHC 
are shown in Fig.~\ref{fig:nevt} as a function of $1/R$.
Out of the total number of events with 5 leptons or more at $1/R=500$ GeV, 
the majority arise from first- and second- 
generation weak doublet quarks, either in pairs or in association with other particles; 
$W_\mu^\one$ pair production is responsible for around 10\%, 
as is production including $SU(3)_c$ bosons, $G_{\mu ,H}^\one$.
As parton distribution functions vary with the size of the extra dimensions, 
so will the individual contributions, 
although the sensitivity to the mass scale $1/R$ is small. 
The results shown in Fig.~\ref{fig:nevt} include tree-level processes only. 
We estimate that next-to-leading order effects will increase the cross sections by $\sim$30-50\%, especially due to initial state radiation.
A complete analysis of this effect is warranted, but is beyond the scope of this paper.

\begin{figure}[b]
\centerline{
\psfig{file=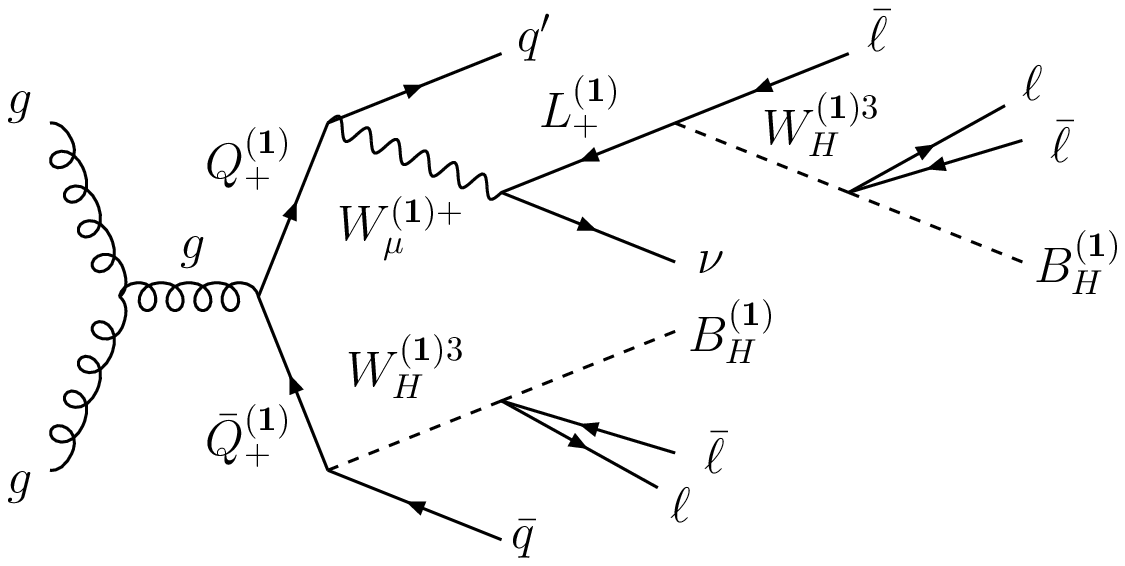,width=7.5cm,angle=0}
\hspace{0.2cm}
\psfig{file=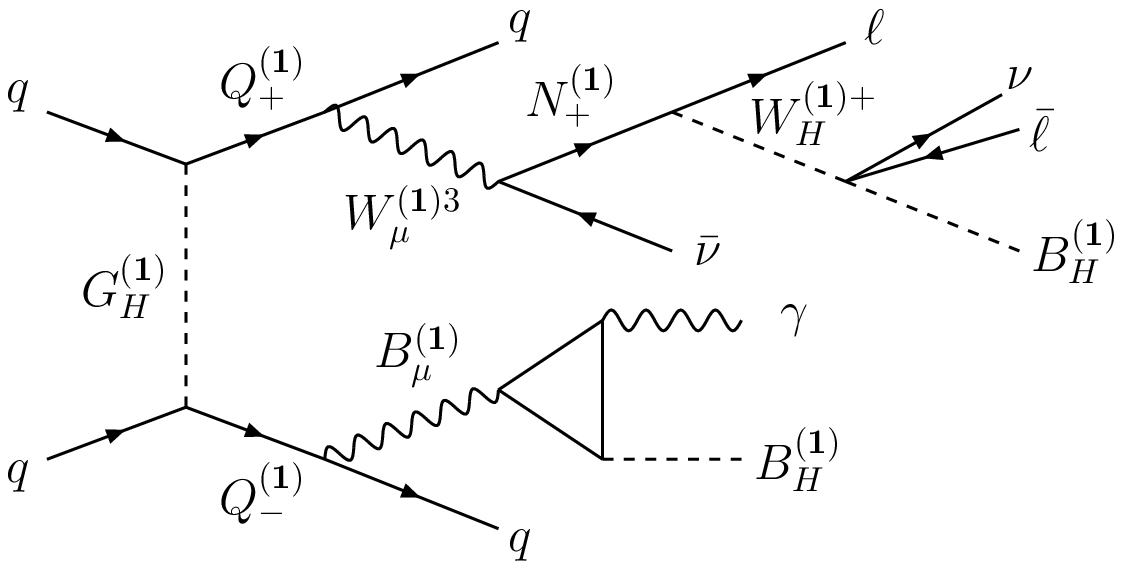,width=7.5cm,angle=0}
}
\vspace*{-1.4mm}
\caption{Representative processes that lead to  
$5\ell + \met$ and $\gamma \ell^+\ell^- +\met$ events.  Several other 
production mechanisms as well as cascade decays contribute to these and related 
signals. 
\label{fig:ex-diagrams}}
\end{figure}

Also interesting are combined photon and lepton events which result from 1-loop decays of 
the $\one$ hypercharge gauge boson $B_\mu^\one$ produced in the decay chain of $U_-^\one$ quarks (see Fig.~\ref{fig:eeee}(a)).
Down-type quarks have smaller hypercharge and so couple less strongly; while quark doublets couple more strongly to weak bosons, resulting in a negligible 
branching fraction into $B_\mu^\one$.  In Fig.~\ref{fig:ex-diagrams} we show typical diagrams for 
$\ell^+ \ell^+ \ell^+ \ell^- \ell^-$ and $\gamma \ell^+ \ell^- $ signatures.
The rate for events with unusual
combinations of final states: two same-sign leptons and a photon,
$\gamma\ell^+ \ell^+$ ($\gamma\ell^- \ell^-$) for instance, or three 
same-sign and one opposite sign lepton, $\ell^+\ell^+\ell^+\ell^-$ 
($\ell^-\ell^-\ell^-\ell^+$), are plotted in Fig.~\ref{fig:eeee}(b).  The
latter process consists of around 10\% of the total rate for 4 lepton events, 
and the largest single contribution to it is the decay of $U_+^\one$ 
($D_+^\one$) pairs. It arises only rarely in the standard model from $W^+W^+Z$ ($W^-W^-Z$) production.
%
%

We expect that the small standard model backgrounds for these processes can be
eliminated by using a hard $\met$ cut in conjunction with a jet $p_T$
cut since the jets originating from the decay of (1,0) colored
particles should have a transverse momentum of the order of their mass
differences ($\sim 100$ GeV). One might also naively worry about
triggering issues due to the softness of leptons, since the cascade
decays giving rise to them occur between particles that are relatively
degenerate in mass. A preliminary analysis on a single leg of the
decay chain keeping exact spin correlations suggests that more than 90 \% of 
lepton pairs have enough $p_T$ to evade a 15 GeV cut, 
 and that the leptons
are far enough away in $\Delta R$ to be visible as individual tracks.
Hence we do not anticipate any triggering problems, although a
detailed analysis of these issues using a detector simulator might be beneficial. 

\subsection{Cross sections at the Tevatron} 
%
%

\begin{figure}[t]
\centerline{
\psfig{file=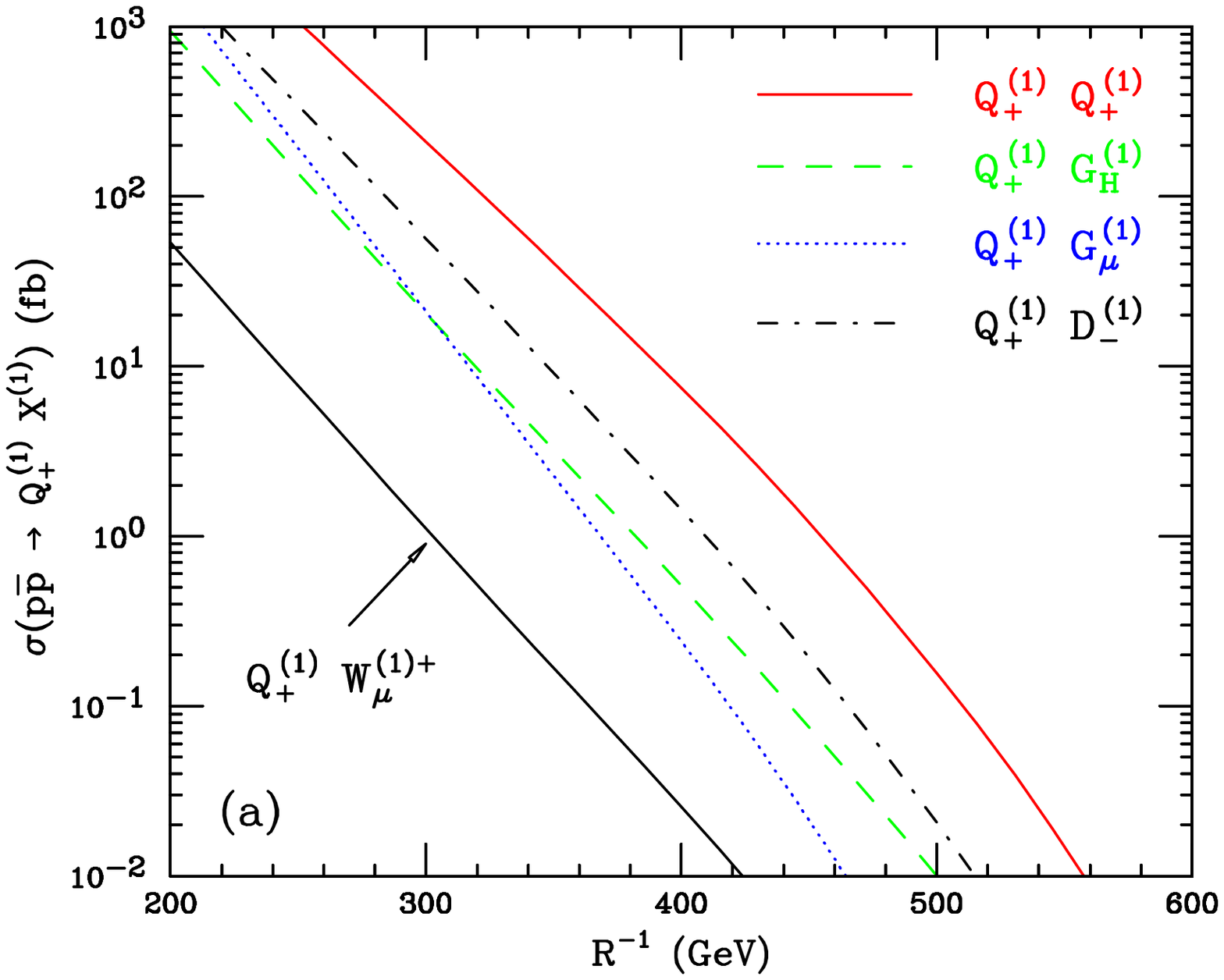,width=8cm,angle=0}
\psfig{file=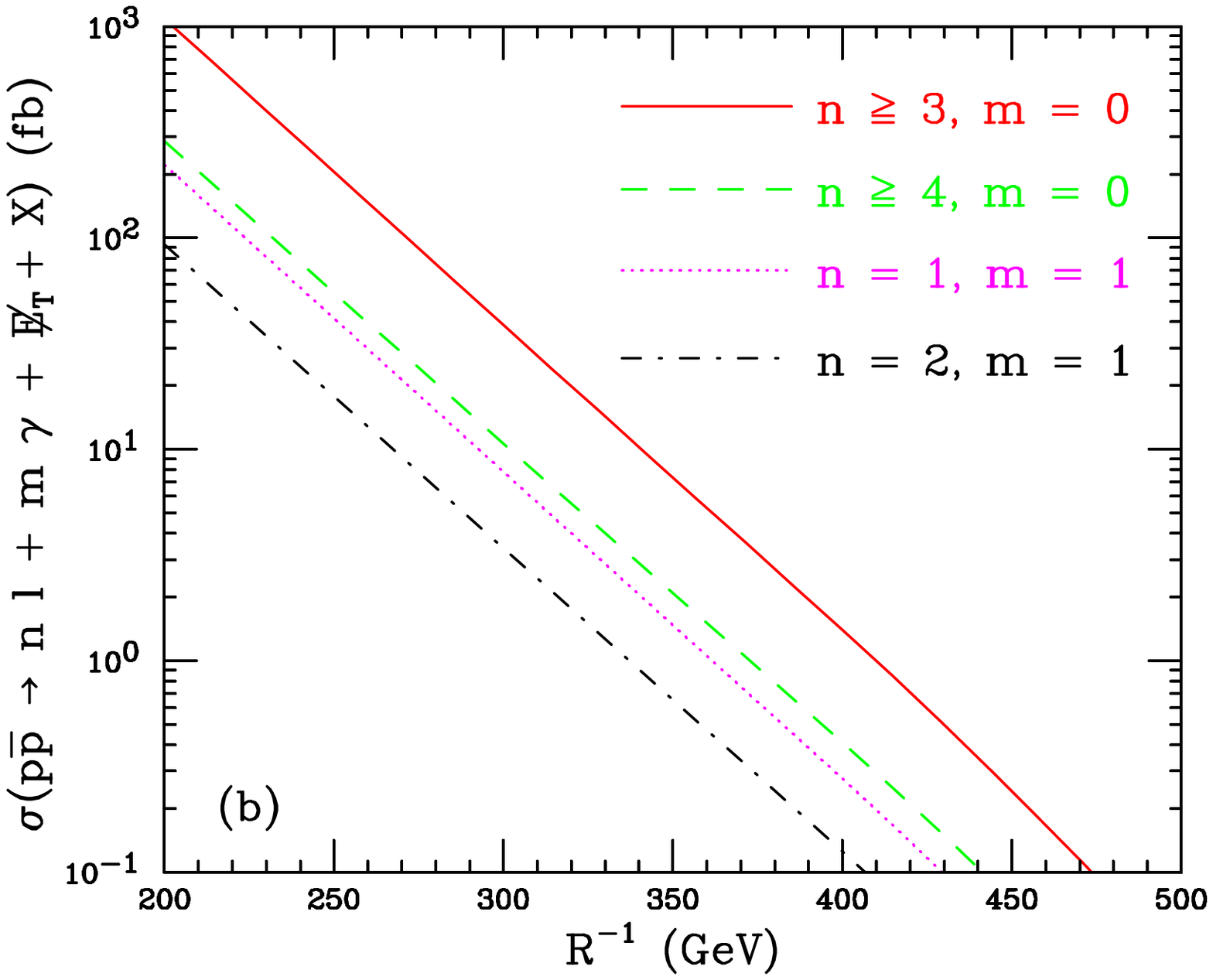,width=8cm,angle=0}}
\vspace*{-4mm}
\caption{(a) Production cross sections at the Tevatron and  (b) Cross sections for multilepton
+ photon events, as a function of $1/R$.}
\label{fig:prod-TeV}
\end{figure}

At the Tevatron, the production from a $q\bar{q}$ initial state,
shown in Figs.~\ref{fig:qqQQ'},~\ref{fig:qqGHGH} and~\ref{fig:qqW+GH}, dominates.
We summarize our results for (1,0) production cross sections, as well as multi-lepton and lepton plus photon signatures in
Fig.~\ref{fig:prod-TeV}.
The lower center-of-mass energy of this collider slightly increases $W_\mu^\one$ production cross sections
as compared with the LHC. This process now contributes 16\% of the total number of events
with 4 or more leptons for $1/R=300$ GeV.

We can use data gathered from Tevatron Run II to place rough constraints
on the radius of the extra dimensions.  One potential channel that has been
searched for in the context of the minimal supersymmetric standard model is the trilepton signal~\cite{CDFnote,D0note}.  We apply the results of this analysis, which found no excess over standard model background, directly to our model.  If we assume an efficiency of $\sim5\%$~\cite{CDFnote,D0note}, we see that $1/R$ must be larger 
than $\sim 270$ GeV, otherwise we might have expected to observe at least 3 events. 
Low statistics for this final state, both in expected and observed events, make the limit rather less reliable than desired.

A more precise, though less stringent, constraint can be obtained by using Run II lepton + photon 
data~\cite{Abulencia:2007zi}, which contains larger numbers of expected and observed events.  The standard model prediction for the $\ell  \gamma  X$ channel
 for instance, is $150.6\pm 13$ with an observation of 163 events. Assuming 
that universal extra dimensions are responsible for the small excesses in this and the $\ell^+ \ell^- \gamma X$ channels allows us to obtain a limit on $1/R$ 
of around 240 GeV at 95\% C.L.



%
%
%
%

\section{Conclusions}
\label{sec:conclusions}
\setcounter{equation}{0}

Despite the successful predictions of the 6DSM, the
hadron collider phenomenology of (1,0) KK modes  has not
been previously studied due to the large number of mechanisms
that contribute to production cross sections. 
Our inclusion in CalcHEP of the interactions between (1,0) particles and
standard model ones has allowed us to compute the 
cross sections for (1,0) pair production at the LHC and the Tevatron.
The large cross sections (of almost $10^4$ fb at the LHC, for masses 
around 500 GeV) shows that cascade decays with small branching fractions
may be observed, leading to a variety of discovery channels.
These are particularly interesting because of the
presence in the 4D effective theory of a spinless adjoint particle for each
standard model gauge group.
One-loop corrections to the level-1 masses tend to make these spinless adjoints lighter
than matter fields~\cite{Ponton:2005kx} 
(the same result~\cite{Azatov:2007fa} applies to other models~\cite{Mohapatra:2002ug}), 
forcing them to undergo tree-level 3-body decays
and emitting two standard model fermions each time.  This results in significant
numbers of events with five or more leptons. 
 
Multi-lepton events are not unique to the 6DSM, although the rates at which they occur in other theories are typically smaller. 
In its 5D counterpart for example, it is 
necessary to produce level-2 KK particles to give rise to long enough
cascades; the rate for such processes is suppressed because the particles
produced are heavier ($m\sim 2/R$) \cite{Datta:2005zs}. 
Another theory leading to multi-lepton signatures involves a warped extra dimension with custodial 
symmetry~\cite{Dennis:2007tv}, but leptons in that case come from decays of $W$ and $Z$, 
whose branching fractions are small.
In supersymmetric models, cascade decays of squarks such as $\tilde{q}'_L \to \tilde{\chi}_2^\pm q 
\to W^\pm \tilde{\chi}_2^0 q (\tilde{\chi}_1^\pm Z q)$ can also give
multi-lepton signatures at the cost of small production cross sections due to spin-statistics 
as well as a small branching fraction for $\tilde{q}'_L \to \tilde{\chi}_2^\pm q$. 

Nevertheless, it should be rather straightforward to differentiate
among these models if a sufficiently large number of multi-lepton events will
be observed at the LHC. The 6DSM has specific preditions for many observables. 
In this paper we analyzed the rates for events with 3, 4, 5 and 6 leptons,
as well as the relative rates for events with three leptons of one charge
and one lepton of opposite charge. Other observables, such as the relative rates
for events with different numbers of electrons and muons, may be analyzed using
the branching fractions for complete cascade decays (see the tables in Section 3).
Another peculiarity of the 6DSM cascade decays is that they lead with reasonably large 
branching fractions to events with photons. This is a consequence of the
2-body decay at one loop of the hypercharge (1,0) vector boson, which competes
successfully with its tree-level 3-body decays.
Events with leptons, photons and missing energy are also predicted in certain 
supersymmetric extensions of the standard model, but again, there are several 
different channels, and we expect that if such events will be seen in large numbers,
it will be possible to differentiate between models.

One may wonder how robust our predictions are against variations in the 
mass spectrum, which may get contributions from operators localized at 
the fixed points of the chiral square, as well as from higher-order QCD effects. 
In the case of a single universal extra dimension,
deviations from the one-loop corrected mass spectrum lead to a variety of 
phenomenological implications \cite{Cembranos:2006gt}.
Within the 6DSM, we expect that the rates for
multi-lepton events remain relatively large when the (1,0) mass spectrum 
is perturbed. 
This is due to the large number of particles involved in a typical decay chain, 
with a standard model quark or lepton being emitted at each stage.
The total rates computed here 
are sums over many such cascade decays of several (1,0) particles.
However, the events with photons depend entirely on the branching fractions of 
a single  particle, the hypercharge vector boson, and thus are less 
generic for different mass spectra.

A more general approach would be to lift the constraints on the mass spectrum.
If excess events with leptons, missing energy and possibly photons 
will be observed in certain channels at the LHC, then the (1,0) masses 
would be determined by comparing a large set of observed rates with the 
6DSM predictions.
One should also keep in mind that the predictions of the 6DSM are not 
limited to collider signals. For example, 
an interesting 
feature is that the LKP has spin 0, with various implications for dark matter
\cite{next}.

\bigskip

{\bf Acknowledgments:} \ We would like to thank Hsin-Chia Cheng, Konstantin Matchev and Eduardo Ponton for
helpful conversations.
Fermilab is operated by Fermi Research Alliance, LLC under Contract No. 
DE-AC02-07CH11359 with the United States Department of Energy. 

\section*{Appendix A: \ Feynman rules for (1,0) modes}
\addcontentsline{toc}{section}{Appendix A: \ Feynman rules for (1,0) modes}
\label{app:diagrams}
\renewcommand{\theequation}{A.\arabic{equation}}
\setcounter{equation}{0}

In this section we show Feynman rules that are relevant 
for QCD production of (1,0) particles at hadron colliders.
Corresponding vertices involving electroweak gauge bosons can be easily inferred from those given below.
The vector-like nature of KK fermions allows for the usual QCD coupling to standard model gluons 
seen in the $G_\mu Q^{\one}\overline{Q}{^\one}$ vertex below.
\begin{figure*}[h]
\unitlength=1.0 pt
\SetScale{1.0}
\SetWidth{1}      
\normalsize   
\large     
\begin{center}
\begin{picture}(350,80)(40,20)
\Gluon(30,50)(70,50){3.2}{4}
\Vertex(70,50){2}
\ArrowLine(70,50)(100,80)
\ArrowLine(100,20)(70,50)
\Text(25,50)[r]{$G_\mu^{a}$}
\Text(130,90)[r]{$Q^\one_{\pm}$}
\Text(130,10)[r]{$Q^\one_{\pm}$}
\Text(130,50)[l]{$ = - i g_s \gamma^\mu T^a$}

\Photon(240,50)(280,50){3}{4}
\Vertex(280,50){2}
\ArrowLine(280,50)(310,80)
\ArrowLine(310,20)(280,50)
\Text(235,50)[r]{$G_\mu^{\one a}$}
\Text(340,90)[r]{$Q^\one_{\pm}$}
\Text(340,10)[r]{$Q^{(0,0)}_{\pm}$}
\Text(340,50)[l]{$ = - i g_s \gamma^\mu P_{\substack{L \\ R}} T^a$ }

\end{picture}
\end{center}
%
\end{figure*}

The interaction of a level-1 quark and a level-1 gluon is chiral and 
so its vertex contains projection operators, 
although the chirality of the incoming fermion is conserved.

\begin{figure*}[h]
\unitlength=1.0 pt
\SetScale{1.0}
\SetWidth{1}      
\normalsize   
\large     
\begin{center}
\begin{picture}(350,80)(40,20)
\DashLine(30,50)(70,50){3}
\Vertex(70,50){2}
\ArrowLine(70,50)(100,80)
\ArrowLine(100,20)(70,50)
\Text(25,50)[r]{$G_H^{\one a}$}
\Text(130,90)[r]{$Q^\one_{\pm}$}
\Text(130,10)[r]{$Q^{(0,0)}\pm$}
\Text(130,50)[l]{$= - g_s P_{\substack{L\\ R}} T^a$}

\Gluon(240,50)(280,50){3.2}{4}
\Vertex(280,50){2}
\DashLine(280,50)(310,80){3}
\DashLine(310,20)(280,50){3}

\put(305,85){\vector(-1,-1){20}}
\put(305,15){\vector(-1,1){20}}
\Text(297,85)[r]{$p$}
\Text(297,15)[r]{$q$}
\Text(235,50)[r]{$G_\mu^{a}$}
\Text(345,90)[r]{$G_H^{\one b}$}
\Text(345,10)[r]{$G_H^{\one c}$}
\Text(340,50)[l]{$ = g_s f^{abc} \big ( p - q \big )^\mu$ }
\end{picture}
\end{center}
%
\end{figure*}

However, the interaction of a spinless adjoint $G_H^{(1,0)a}$ with fermions changes the chirality of 
the incoming fermion since $G_H^{(1,0)a}$ is a scalar. 
Note that the Feynman rules for standard-model gluons are fixed by gauge invariance.
The 3 and 4-point interactions involving only (1,0) vector bosons and zero-mode gluons
are identical to those in the standard model.

%
%
\begin{figure*}[h]
\unitlength=1.0 pt
\SetScale{1.0}
\SetWidth{1}      
\normalsize   
\large     
\begin{center}
\begin{picture}(350,80)(40,20)
\Vertex(70,50){2}
\Gluon(70,50)(35,80){3.2}{4}
\Gluon(35,20)(70,50){3.2}{4}
\DashLine(70,50)(100,78){3}
\DashLine(100,22)(70,50){3}
\Text(35,86)[r]{$G_\mu^{b}$}
\Text(35,14)[r]{$G_\nu^{d}$}
\Text(135,86)[r]{$G_H^{\one c}$}
\Text(135,14)[r]{$G_H^{\one e}$}
\Text(130,50)[l]{$ = -i g_s^2 g^{\mu\nu} (f^{abc} f^{ade} + f^{abe}f^{adc})$}
\end{picture}
\end{center}
\vspace*{.3cm}
\end{figure*}
\vspace{0.2cm}

%
%
\begin{figure*}[h!]
\unitlength=1.0 pt
\SetScale{1.0}
\SetWidth{1}      
\normalsize   
\large     
\begin{center}
\begin{picture}(350,80)(40,20)
\Gluon(30,50)(70,50){3.2}{4}
\Vertex(70,50){2}
\Photon(70,50)(100,80){3}{4}
\Photon(100,20)(70,50){3}{4}
\Text(25,50)[r]{$G_\nu^{b}$}
\Text(130,90)[r]{$G_\mu^{\one a}$}
\Text(130,10)[r]{$G_\rho^{\one c}$}
\Text(130,50)[l]{$= g_s f^{abc} \big [ (k-p)_\lambda g_{\mu\nu} + 
(p-q)_\mu g_{\nu\rho} + (q-k)_\nu g_{\mu\rho} \big ] $}
\put(88,83){\vector(-1,-1){20}}
\put(106,28){\vector(-1,1){20}}
\put(35,40){\vector(1,0){30}}
\Text(77,77)[r]{$k$}
\Text(105,45)[r]{$q$}
\Text(55,30)[r]{$p$}
\end{picture}
\end{center}
\end{figure*}
\vspace{0.2cm}
\begin{figure*}[h!]
\unitlength=1.0 pt
\SetScale{1.0}
\SetWidth{1}      
\normalsize   
\large     
\begin{center}
\begin{picture}(350,80)(40,20)
\Vertex(69,50){2}
\Gluon(70,50)(38,77){3.2}{4}
\Gluon(38,23)(70,50){3.2}{4}
\Photon(70,50)(97,77){3}{3}
\Photon(97,23)(70,50){-3}{3}
\Text(38,84)[r]{$G_\mu^{a}$}
\Text(38,12)[r]{$G_\nu^{b}$}
\Text(131,84)[r]{$G_\rho^{\one c}$}
\Text(131,14)[r]{$G_\sigma^{\one d}$}
\Text(130,70)[l]{$ = -i g_s^2 \big [ f^{abe} f^{cde} (g^{\mu\rho} g^{\nu\sigma} - g^{\mu\sigma} g^{\nu\rho} ) + f^{ace} f^{bde} (g^{\mu\nu} g^{\rho\sigma} - g^{\mu\sigma} g^{\nu\rho} ) $ }
\Text(150,40)[l]{$+ f^{ade} f^{bce} (g^{\mu\nu} g^{\rho\sigma} - g^{\mu\rho} g^{\nu\sigma} ) \big ] $ }
\end{picture}
\end{center}
\end{figure*}

%
%
%

\vspace{0.35cm}
\section*{Appendix B: \ One-loop 2-body decays of (1,0) bosons}
\addcontentsline{toc}{section}{Appendix B: \ One-loop 2-body decays of (1,0) bosons}
\renewcommand{\theequation}{B.\arabic{equation}}
\setcounter{equation}{0}
We compute here the 
amplitude  for the process 
$B_\nu^{\one} \rightarrow B_H^{\one} \gamma $, which proceeds through 
one-loop diagrams with KK fermions running in the loop.
The couplings of the $B_\nu^{\one}$ and $B_H^{\one}$ bosons to the KK modes of a 
6D chiral fermion $F_+$ are given by
\begin{eqnarray} \hspace*{-2cm}
\mathcal{L} \supset \frac{1}{4} g^\prime Y_{F_+} \overline{F}_+^{(j,k)} \hspace*{-0.2cm}
&& \left[ B_\nu^{\one} \gamma^\nu \left( P_L \, d_{00}^{j,k;j^\prime\!,k^\prime} 
- P_R \, d_{10}^{j,k;j^\prime\!,k^\prime}r_{jk}^*r_{j^\prime,k^\prime}\right) \right.
\nonumber \\ [2mm]
&& \left.  \
- \, i B_H^{\one}\left( P_R \, d_{01}^{j,k;j^\prime\!,k^\prime} r_{j^\prime,k^\prime}
- P_L \, d_{03}^{j^\prime \!,k^\prime\!;j,k}r_{jk}^*\right) \right] F_+^{j^\prime,k^\prime}
~.
\end{eqnarray}
Here we have defined 
\begin{eqnarray}
d_{n n^\prime}^{j,k;j^\prime\!,k^\prime} & = & (-1)^{n} \delta_{k^\prime \!,k}
\left(\delta_{j^\prime \!, j-1} +   (-1)^{n^\prime}\delta_{j^\prime \!, j+1} \right) 
+ (-1)^{n} \delta_{j^\prime \!,j}\left( \delta_{k^\prime \!, k+1} 
+ (-1)^{n^\prime}\delta_{k^\prime \!, k-1} \right) 
\nonumber \\ [2mm]
&& + \, i^{n^\prime - n} \delta_{j,1}\delta_{k^\prime \!,0}\delta_{j^\prime \!,k}
+ i^{n+ 2n^\prime} \delta_{j^\prime \!,1}\delta_{k,0}\delta_{k^\prime \!,j} ~,
\end{eqnarray}
where $r_{j,k}$ are complex phases, 
\be
r_{j,k} = \frac{j+i k}{\sqrt{j^2+k^2}}
\ee
and $Y_F$ is the hypercharge of the fermion, normalized to $-1$ for lepton doublets.
In the case of fermions with 6D  chirality $-$, which contain right-handed zero modes,
the same formulas apply with the 
$P_L$ and $P_R$ chirality projection operators interchanged.

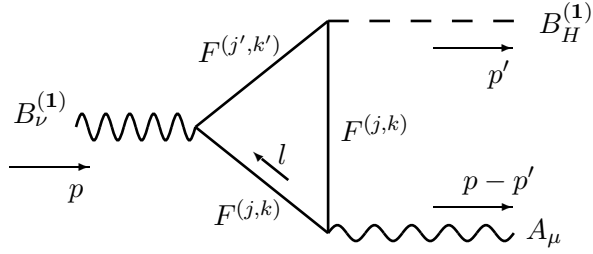
\begin{figure}[t]
\unitlength=1.0 pt
\SetScale{1.0}
\SetWidth{1.}      
\normalsize     
\begin{center}
\begin{picture}(110,80)(40,20)
\Photon(5,40)(50,40){5}{5}
\Photon(100,0)(170,0){3}{5}
\DashLine(100,80)(170,80){8}
\Line(50,40)(100,80)
\Line(100,0)(50,40)
\Line(100,0)(100,80)
\Text(2,47)[r]{$B_\nu^{\one}$}
\Text(180,80)[l]{$B_H^{\one}$}
\Text(182,0)[c]{$A_\mu$}
\Text(83,70)[r]{$ F^{(j^\prime,k^\prime)}$}
\Text(83,8)[r]{$ F^{(j,k)}$}
\Text(105,40)[l]{$ F^{(j,k)}$}
\put(-20,25){\vector(1,0){30}}
\Text(5,15)[c]{$p$}
\put(140,10){\vector(1,0){30}}
\Text(165,20)[c]{$p-p^\prime$}
\put(140,70){\vector(1,0){30}}
\Text(165,60)[c]{$p^\prime$}
\put(76,27){\vector(-1,1){4}}
\Line(85,20)(75,28)
\Text(83,30)[c]{$l$}
\end{picture}
\end{center}
\vspace{0.5cm}
\caption{Dimension-5 operator induced by fermion loops.}
\label{fig:one-loop}
\end{figure}


Dimension-5 operators coupling a (1,0) vector boson to a (1,0) 
spinless adjoint and a standard-model gauge boson are induced 
at one loop by the diagram in Figure \ref{fig:one-loop}, with fermion KK modes running 
in the loop.
The contribution of a fermion $F_+$ to the 
amplitude for $B_\nu^{\one} \rightarrow B_H^{\one} \gamma_\mu$
is given by
\be
{\cal M}\left( B_\nu^\one \rightarrow B_H^\one \gamma_\mu \right)_{F_+}  =
- \frac{1}{4} \left(g^{\prime} \frac{Y_{F_+}}{2} \right)^{\! 2} e\, Q_{F_+}\,
\varepsilon_\mu^*(p-p^\prime) \, \varepsilon_\nu (p) \, I_{F_+}^{\mu\nu (j,k;j^\prime \!, k^\prime)}  ~,
\ee
where 
\be 
I_{F_+}^{\mu\nu (j,k;j^\prime \!,k^\prime)} = \int \!\!\frac{d^4 l}{(2\pi)^4}
{\rm Tr}
\frac{m_F^{j,k;j^\prime \!, k^\prime} 
\left[\slal \gamma^\mu + \gamma^\mu ( \slal + \slap - \slapp )\right] 
(\slal + \slap) 
-  m_F^{j^\prime \!, k^\prime \!;j,k} \slal\gamma^\mu 
( \slal + \slap - \slapp) }
{\left(l^2 - M^2_{F^{(j,k)}}\right)\left[(l+p-p^\prime)^2 - M^2_{F^{(j,k)}}\right]
\left[(l+p)^2 - M^2_{F^{(j^\prime,k^\prime)}}\right]}  \gamma^\nu\gamma_5
\ee
and
\be
m^{j,k;j^\prime \!, k^\prime}_F  = M_{F^{(j,k)}} \, {\rm Re} \left[r_{jk}  
\left(d_{00}^{j,k;j^\prime \!, k^\prime}
d_{01}^{j^\prime\!,k^\prime\!;j, k}
- d_{10}^{j^\prime\!,k^\prime\!;j, k}
d_{01}^{j,k;j^\prime\!, k^\prime} \right)\right] ~.
\ee
After integrating over the loop momentum $l$, and summing over fermions, we find the 
amplitude 
\be\label{equ:3bodydecay}
{\cal M}\left( B_\nu^\one \rightarrow B_H^\one \gamma_\mu \right)  = 
- \frac{g^{\prime 2} e}{8 \pi^2} \epsilon^{\mu\nu\alpha\beta}
\frac{\varepsilon_\mu^*(p-p^\prime) \varepsilon_\nu (p) p_\alpha p^\prime_\beta }{M_{B_\nu^\one}^2 
- M_{B_H^\one}^2} \sum_F  \sigma_F \, \left(\frac{Y_F}{2}\right)^2 \, Q_F \, {\cal E}_F ~,
\ee
where $\sigma_F = \pm 1$ when $F$ has 6D chirality $\pm$, and 
\be
{\cal E}_F = \sum_{j,k;j^\prime\!, k^\prime}
 m_F^{j,k;j^\prime \!, k^\prime} J_F^{j,k;j^\prime\!, k^\prime} ~,
\ee
with $J_F$ given by an integral over a Feynman parameter:
\be
J_F^{j,k;j^\prime\!, k^\prime} = \int_0^1 \frac{dx }{x} 
\ln \left( 1 + \frac{x(1-x) 
\left(M_{B_\nu^\one}^2 - M_{B_H^\one}^2\right)}
{(1-x)M_{F^{(j,k)}}^2 + x M_{F^{(j^\prime\!,k^\prime)}}^2 - x(1-x) M_{B_\nu^\one}^2} \right) ~.
\ee

The $m^{j,k;j^\prime \!, k^\prime}$ quantities vanish unless the set of KK numbers
$(j,k;j^\prime \!, k^\prime)$ is given by (1,0;1,1), (1,1;1,0) or (1,0; 0,0).
This is a consequence of the vectorlike nature of the fermion higher KK modes.
Therefore,
\be
{\cal E}_F 
= M_{F^{(1,0)}}\left( 2 J_F^{1,0; 0,0} + J_F^{1,0;1,1} \right)
+ \sqrt{2} M_{F^{(1,1)}} J_F^{1,1;1,0} ~.
\label{ef}
\ee
Note that ${\cal E}_F$ depends only on the (1,0) masses and on the masses of the (0,0) and 
(1,1) fermions.  The mass corrections for (1,1) fermions, 
$\big\{Q^3_+, T_-,Q^{1,2}_+,U^{1,2}_-,D^{1,2,3}_-,L_+$ and $E_-\big\}$,
are given by $\sqrt{2}/R$ multiplied by the coefficients 
$\left\{1.33,1.31,1.31,1.27,1.26,
1.05,1.02\right\}$ respectively \cite{Burdman:2006gy},
ignoring electroweak symmetry breaking effects.
 Note also that in the limit that all the fermions at each
KK level are degenerate, ${\cal E}_F$ becomes independent of $F$ and so can be taken out of the sum in
Eq.~(\ref{equ:3bodydecay}), which then vanishes identically by anomaly cancellation.  
This completes the computation of the amplitude  for 
$B_\nu^{\one} \rightarrow B_H^{\one} \gamma $, which determines the coefficient of 
the dimension-5 operator shown in Eq.~(\ref{operator}), and the decay width of 
$B_\nu^{\one}$ shown in Eq.~(\ref{oneloopdecay}). 

\bigskip

\section*{Appendix C: \ Tree-level 3-body decays of (1,0) bosons}
\addcontentsline{toc}{section}{Appendix C: \ Tree-level 3-body decays of (1,0) bosons}
\label{app:3body}
\renewcommand{\theequation}{C.\arabic{equation}}
\setcounter{equation}{0}

\begin{figure*}[t]
\begin{center}
\unitlength=1.0 pt
\SetScale{1.0}
\SetWidth{1}      
\begin{picture}(360,90)(25,30)
\DashLine(50.0,72.5)(90.0,72.5){3}
\Photon(50.0,68.0)(90.0,68.0){2.5}{6}
\Text(95.0,70.0)[r]{$\bullet$}
\ArrowLine(90.0,70.0)(130.0,90.0)
\ArrowLine(130.0,50.0)( 90.0,70.0)
\ArrowLine(170.0,30.0)(130.0,50.0)
\DashLine(170.0,72.0)(130.0,51.0){3}
\Photon(169.0,67)(133.0,48.0){2.5}{6}
\Text(134.0,49.0)[r]{$\bullet$}
\Text( 45.0,73.0)[r]{$A_2$}
\Text(190.0,70.0)[r]{$A_1$}
\Text(110.0,50.0)[r]{$F$}
\Text(140.0,95.0)[r]{$f$}
\Text(180.0,25.0)[r]{$\bar{f}$}
\Text(  210,50)[r]{+}
\DashLine(230.0,72.5)(270.0,72.5){3}
\Photon(230.0,68.0)(270.0,68.0){2.5}{6}
\Text(275.0,70.0)[r]{$\bullet$}
\ArrowLine(270.0,70.0)(310.0,90.0)
\ArrowLine(310.0,90.0)(350.0,110.0)
\ArrowLine(310.0,50.0)(270.0,70.0)
\DashLine(350.0,73.5)(315.0,91.0){3}
\Photon(350.5,69.0)(310.0,89.0){2.5}{6}
\Text(316.0,90.0)[r]{$\bullet$}
\Text(225.0,73.0)[r]{$A_2$}
\Text(365.0,110.0)[r]{$f$}
\Text(370.0,65.0)[r]{$A_1$}
\Text(297.0,90.0)[r]{$F$}
\Text(320.0,50.0)[r]{$\bar{f}$}
\end{picture}
\end{center}
\caption{The diagrams for 3-body decay of (1,0) particles. 
$A_2$ and $A_1$ are heavy bosons of spin 0 or 1, $F$ is a heavier fermion,
and $f$ is a much lighter fermion.}
\label{fig:diagrams}
\end{figure*}
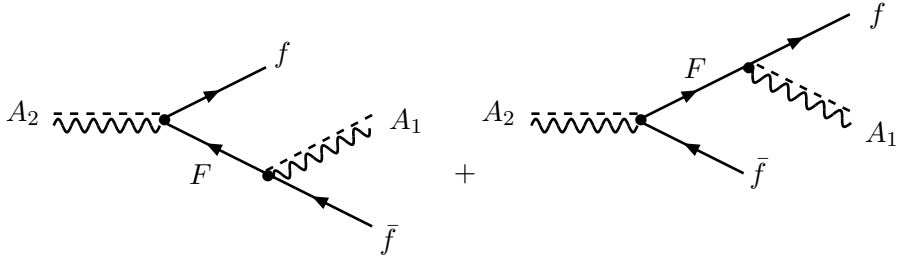

In this Appendix we compute the width for 3-body decays of (1,0) bosons.
Let us consider a generic 3-body decay of a boson
$A_2$ of mass $M_2$ into a boson $A_1$ of mass $M_1$
and a fermion-antifermion pair $f \bar{f}$, 
via an off-shell fermion $F$, of mass $M_F > M_2 > M_1$. 
There are two tree-level diagrams contributing to the process
$A_2 \to (F^\ast f) \to A_1 f \bar{f}$, as shown in Fig.~\ref{fig:diagrams}.
For simplicity, we assume that the final-state fermions are massless.
The decay width is given by
\begin{equation}
\Gamma (A_2 \to A_1 f \bar{f})
= \frac{1}{64\pi^3 M_2} 
\int_{0}^{\e} dE_f \int_{\e-E_f}^{E_{\bar{f}}^{\rm max}} dE_{\bar{f}} \,
\, \overline{ \left|\cal{M}\right|^2 } ~,
\end{equation}
where $\cal{M}$ is the matrix element, $E_f$ and $E_{\bar{f}}$ are the 
energies of the final-state fermions in the rest frame of $A_2$, and we defined
\be
\e \equiv \frac{M_2^2 - M_1^2}{2 M_2} \, .
\label{mu21}
\ee
For a fixed $E_f$, the maximum value of $E_{\bar{f}}$ is 
\begin{eqnarray}
E_{\bar{f}}^{\rm max} &=& \frac{ \e - E_f }{ 1- 2E_f/M_2} \, .
\label{Emax}
\end{eqnarray}
%
%

Let us first consider the case where both $A_1$ and $A_2$ have spin 0
(we label them by $A_{1H}$ and $A_{2H}$ in that case) and have pseudo-scalars
couplings to the fermions: 
\be 
\left( g_1 A_{1 H} + g_2 A_{2 H} \right) i \overline{F}_L f_R + {\rm H.c.} ~,
\label{couplings-ps}
\ee
where $g_{1,2}$ are real dimensionless couplings.
The matrix element squared, summed over the spins of $f$ and $\bar{f}$, is given by
\be
\overline{\left|{\cal M}\right|^2} \left( A_{2H} \to f_R \bar{f}_R A_{1H} \right)
= 2\left( g_1 g_2 \right)^2 
        \Big [ 2 (P_f \cdot P_1)(P_f \cdot P_1)  
              - M_2^2 (P_f \cdot P_{\bar{f}}) \Big ] \, \Delta^2 ~,
\ee
where $P_1$, $P_f$ and $P_{\bar{f}}$ are the 4-momenta of $A_{1H}$, $f$ 
and  $\bar{f}$, respectively. The quantity 
\be
\Delta = \frac{1}{ (P_1 + P_f)^2 - M_F^2} - 
    \frac{1}{ (P_1 + P_{\bar{f}})^2 - M_F^2}  \, ,
\label{Delta}
\ee
accounts for the propagators of the off-shell fermion in the two diagrams 
of Fig.~\ref{fig:diagrams}. 
The two diagrams have opposite sign, resulting in the sign between the two terms
in $\Delta$, because of the different momentum flow through the intermediate fermion line.
In the center-of-mass frame, the width becomes
\begin{equation}
\Gamma (A_{2H} \to A_{1H} f_R \bar{f}_R)
= \frac{\left( g_1 g_2 \right)^2 }{128\pi^3} M_2 \, {\cal I}_+(M_2, M_1, M_F)
\label{AHtoAH}
\end{equation}
where we defined
\begin{equation}
{\cal I}_\pm(M_2, M_1, M_F) =\int_{0}^{\e} dE_f \int_{\e-E_f}^{E_{\bar{f}}^{\rm max}} 
\! dE_{\bar{f}} \; \frac{2E_f E_{\bar{f}} \pm M_2 \left(\e - E_f-E_{\bar{f}}\right) }
         {M_2^2(\mup + E_f)^2 ( \mup + E_{\bar{f}})^2}  \left(E_f-E_{\bar{f}}\right)^2 ~.
\label{fpm}
\end{equation}
The function ${\cal I}_-$ is introduced for later convenience, 
$\e$ and $E_{\bar{f}}^{\rm max}$ are given in Eqs.~(\ref{mu21}) and (\ref{Emax}), respectively, 
and 
\be
\mup \equiv \frac{M_F^2 - M_2^2}{2 M_2} ~~ .
\ee

Let us now study the case where $A_2$ has spin 1
(we label it by $A_{2\mu}$ in that case) and 
couples to one chirality of the fermions: 
\be 
g_2 A_{2 \mu} \overline{F}_R \gamma^\mu f_R + {\rm H.c.} ~.
\ee
The matrix element squared, averaged over the polarizations of $A_{2 \mu}$ 
and summed over the spins of $f$ and $\bar{f}$, is given by
\be
\overline{\left|{\cal M}\right|^2} \left( A_{2\mu} \to f_R \bar{f}_R A_{1H} \right)
= \frac{2}{3} \left( g_1 g_2 \right)^2 
\left(\frac{M_F}{M_2}\right)^{\! 2}  \Big [ 2 (P_f \cdot P_2)( P_{\bar{f}} \cdot P_2)
 + M_2^2 P_f \cdot P_{\bar{f}} \Big ] \Delta^2 ~,
\ee
where $P_2$ is the 4-momentum of $A_{2H}$.
Again, the two diagrams have opposite signs, resulting in the form of $\Delta$
given in Eq.~(\ref{Delta}).
However, the sign difference in this case is due to the pseudo-scalar coupling.
The width in the center-of-mass frame is given by
\begin{equation}
\Gamma (A_{2\mu} \to A_{1H} f_R \bar{f}_R)
= \frac{\left( g_1 g_2 \right)^2 }{384\pi^3} \frac{M_F^2}{M_2} \, {\cal I}_-(M_2, M_1, M_F) ~,
\label{AmutoAH}
\end{equation}
where ${\cal I}_-$ is the phase-space integral shown in Eq.~(\ref{fpm}).

The only other case relevant for the decays of the (1,0) particles 
discussed in Section~\ref{sec:decays} is that where $A_2$ has spin 0 and pseudo-scalar couplings 
[see Eq.~(\ref{couplings-ps})], while $A_1$ has spin 1 and a coupling
\be 
g_1 A_{1 \mu} \overline{F}_R \gamma^\mu f_R + {\rm H.c.} ~.
\ee
The matrix element squared, summed over the polarizations of $A_{1 \mu}$ 
and the spins of $f$ and $\bar{f}$, is given in this case by
\be
\overline{\left|{\cal M}\right|^2} \left( A_{2H} \to f_R \bar{f}_R A_{1\mu} \right)
= 2 \left( g_1 g_2 \right)^2 
\left(\frac{M_F}{M_1}\right)^{\! 2}  \Big [ 2 (P_f \cdot P_1)( P_{\bar{f}} \cdot P_1)
                            + M_2^2 P_f \cdot P_{\bar{f}} \Big ] \Delta^2 ~,
\ee
where $\Delta$ is defined in Eq.~(\ref{Delta}).
The width in the center-of-mass frame is given by
\begin{equation}
\Gamma (A_{2H} \to A_{1\mu} {\cal I}_R \bar{f}_R)
=  \frac{\left( g_1 g_2 \right)^2 }{128\pi^3} M_2 \frac{M_F^2}{M_1^2} \, 
\left[ \left(1 - \frac{2\e}{M_2}\right) {\cal I}_-(M_2, M_1, M_F) 
+ \frac{2\e}{M_2}{\cal I}_+(M_2, M_1, M_F) \right] ~.
\label{AHtoAmu}
\end{equation}

If the heavy particles are approximately degenerate, 
which is the case for the (1,0) particles studied in this paper, then
$\e \ll M_2$ and  $\mup \ll M_2$ (which implies $\e\approx M_2-M_1$ and $\mup\approx M_F-M_2$),
and the double integrals of Eq.~(\ref{fpm}) may be performed analytically: 
\bear
\hspace*{-7em}
{\cal I}_+ (M_2, M_1, M_F) & = & 
\frac{-8}{M_2^3}\left[ \rule{0mm}{5mm} \mup \frac{\e + \mup}{\e + 2 \mup}
\left( \e^2 + 5 \e \mup + 5 \mup^2 \right)
\ln \left( 1 + \frac{\e}{\mup}\right) \right.
\nonumber \\ [0.7em]
&& \; \; \; - \left. 
\frac{\e}{12} \left( \e^2 +30 \e \mup + 30 \mup^2 \right) \rule{0mm}{5mm}\right] 
\left[1+ O\left(\frac{\e}{M_2} , \frac{\mup}{M_2}  \right) \right]~.
\eear
A simple relation between the ${\cal I}_\pm$ functions holds at leading order
in $1/M_2$:
\be
{\cal I}_- = 3 {\cal I}_+ \left[1+ O\left(\frac{\e}{M_2} , \frac{\mup}{M_2}  \right) \right]~.
\ee
It is also useful to note that for $\e \ll M_2$ 
{\it and}  $\e \ll \mup$, 
\bear
{\cal I}_+(M_2, M_1, M_F) & = & \frac{\e^7}{105 \, M_2^3 \mup^4} 
\left[1 -2 \frac{\e}{\mup} + \frac{\e}{M_2} 
+ O\left(\frac{\e^2}{\mup^2}, \, \frac{\e^2}{M_2^2} \right)\right] ~,
\nonumber \\ [0.7em]
{\cal I}_-(M_2, M_1, M_F) & = & \frac{\e^7}{35 \, M_2^3 \mup^4} 
\left[1 -2 \frac{\e}{\mup} + \frac{5\e}{3M_2} 
+ O\left(\frac{\e^2}{\mup^2}, \, \frac{\e^2}{M_2^2} \right)\right] ~.
\label{fpe}
\eear
This very strong dependence on $\e \approx M_2-M_1$ 
is somewhat surprising. The phase-space integrals of Eq.~(\ref{fpm}) give three powers of $\e$,
and the matrix element squared appears at first sight to give only one more power 
of $\e$. However, the relative sign of the two diagrams forces a cancellation 
of the leading term within $\Delta$ [see Eq.~(\ref{Delta})], so that $\Delta^2$
gives the $\left(E_f-E_{\bar{f}}\right)^2$ factor in Eq.~(\ref{fpm}),
which accounts for two more powers of  $\e$. 
Furthermore, the integration over $E_{\bar{f}}$ cancels 
the leading term in the $\e$ expansion of the numerator of ${\cal I}_\pm$.
The resulting dependence on the 7th power of $\e$ implies that 
the decay width is extremely suppressed, if $A_2$ and $A_1$ are more
degenerate than the $F-A_2$ pair.


The decay widths given in Eqs.~(\ref{AHtoAH}) and (\ref{AHtoAmu}) are used in 
Section~\ref{sec:decays} for computing the branching fractions of the 
spinless adjoints, while the decay width of Eqs.~(\ref{AmutoAH}) determines 
the branching fractions of the (1,0) hypercharge vector boson.

 \vfil 
\begin{thebibliography}{99} \frenchspacing


\bibitem{Appelquist:2000nn}
  T.~Appelquist, H.~C.~Cheng and B.~A.~Dobrescu,
  ``Bounds on universal extra dimensions,''
  Phys.\ Rev.\  D {\bf 64}, 035002 (2001)
  [arXiv:hep-ph/0012100].

\bibitem{Cheng:2003ju}
  H.~C.~Cheng and I.~Low,
  ``TeV symmetry and the little hierarchy problem,''
  JHEP {\bf 0309}, 051 (2003)
  [arXiv:hep-ph/0308199].

\bibitem{Cheng:2002ab}
  H.~C.~Cheng, K.~T.~Matchev and M.~Schmaltz,
  ``Bosonic supersymmetry? Getting fooled at the LHC,''
  Phys.\ Rev.\  D {\bf 66}, 056006 (2002)
  [arXiv:hep-ph/0205314].

\bibitem{Datta:2005zs}
  A.~Datta, K.~Kong and K.~T.~Matchev,
  ``Discrimination of supersymmetry and universal extra dimensions at  hadron
  colliders,''
  Phys.\ Rev.\ D {\bf 72}, 096006 (2005)
  [Erratum-ibid.\ D {\bf 72}, 119901 (2005)]
  [arXiv:hep-ph/0509246].

\bibitem{Burdman:2006gy}
  G.~Burdman, B.~A.~Dobrescu and E.~Ponton,
  ``Resonances from two universal extra dimensions,''
  Phys.\ Rev.\  D {\bf 74}, 075008 (2006)
  [arXiv:hep-ph/0601186]. 

\bibitem{Burdman:2005sr}
  G.~Burdman, B.~A.~Dobrescu and E.~Ponton,
  ``Six-dimensional gauge theory on the chiral square,''
  JHEP {\bf 0602}, 033 (2006)
  [arXiv:hep-ph/0506334]. \\

\bibitem{Dobrescu:2004zi}
B.~A.~Dobrescu and E.~Pont\'{o}n,
``Chiral compactification on a square,''
JHEP {\bf 0403}, 071 (2004)
[arXiv:hep-th/0401032].

\bibitem{Hashimoto:2004xz}
  M.~Hashimoto and D.~K.~Hong,
  ``Topcolor breaking through boundary conditions,''
  Phys.\ Rev.\  D {\bf 71}, 056004 (2005)
  [arXiv:hep-ph/0409223].


\bibitem{Dobrescu:2001ae}
  B.~A.~Dobrescu and E.~Poppitz,
  ``Number of fermion generations derived from anomaly cancellation,''
  Phys.\ Rev.\ Lett.\  {\bf 87}, 031801 (2001)
  [arXiv:hep-ph/0102010].

\bibitem{Appelquist:2001mj}
  T.~Appelquist, B.~A.~Dobrescu, E.~Ponton and H.~U.~Yee,
  ``Proton stability in six dimensions,''
  Phys.\ Rev.\ Lett.\  {\bf 87}, 181802 (2001)
  [arXiv:hep-ph/0107056].

\bibitem{Hooper:2007qk}
For reviews, see:\\
  D.~Hooper and S.~Profumo,
  ``Dark matter and collider phenomenology of universal extra dimensions,''
  arXiv:hep-ph/0701197. \\
  G.~D.~Kribs,
  ``Phenomenology of extra dimensions,''
  arXiv:hep-ph/0605325.


\bibitem{Cheng:2002iz}
  H.~C.~Cheng, K.~T.~Matchev and M.~Schmaltz,
  ``Radiative corrections to Kaluza-Klein masses,''
  Phys.\ Rev.\  D {\bf 66}, 036005 (2002)
  [arXiv:hep-ph/0204342].

\bibitem{Servant:2002aq}
  G.~Servant and T.~M.~P.~Tait,
  ``Is the lightest Kaluza-Klein particle a viable dark matter candidate?,''
  Nucl.\ Phys.\  B {\bf 650}, 391 (2003)
  [arXiv:hep-ph/0206071]. \\
  H.~C.~Cheng, J.~L.~Feng and K.~T.~Matchev,
  ``Kaluza-Klein dark matter,''
  Phys.\ Rev.\ Lett.\  {\bf 89}, 211301 (2002)
  [arXiv:hep-ph/0207125].

\bibitem{Ponton:2005kx}
  E.~Ponton and L.~Wang,
  ``Radiative effects on the chiral square,''
  JHEP {\bf 0611}, 018 (2006)
  [arXiv:hep-ph/0512304].

\bibitem{Chacko:2005pe}
  Z.~Chacko, H.~S.~Goh and R.~Harnik,
  ``The twin Higgs: Natural electroweak breaking from mirror symmetry,''
  Phys.\ Rev.\ Lett.\  {\bf 96}, 231802 (2006)
  [arXiv:hep-ph/0506256].

\bibitem{Burdman:2006jj}
  G.~Burdman and A.~G.~Dias,
  ``The little hierarchy in universal extra dimensions,''
  JHEP {\bf 0701}, 041 (2007)
  [arXiv:hep-ph/0609181].\\
See also
  G.~Burdman,
  ``Two universal extra dimensions,''
  arXiv:hep-ph/0611064.

\bibitem{Macesanu:2002db}
C.~Macesanu, C.~D.~McMullen and S.~Nandi,
  ``Collider implications of universal extra dimensions,''
  Phys.\ Rev.\  D {\bf 66}, 015009 (2002)
  [arXiv:hep-ph/0201300].\\
  J.~M.~Smillie and B.~R.~Webber,
  ``Distinguishing spins in supersymmetric and universal extra dimension
  models at the Large Hadron Collider,''
  JHEP {\bf 0510}, 069 (2005)
  [arXiv:hep-ph/0507170].\\
  T.~G.~Rizzo,
  ``Probes of universal extra dimensions at colliders,''
  Phys.\ Rev.\  D {\bf 64}, 095010 (2001)
  [arXiv:hep-ph/0106336]. \\
  C.~Macesanu,
  ``The phenomenology of universal extra dimensions at hadron colliders,''
  Int.\ J.\ Mod.\ Phys.\  A {\bf 21}, 2259 (2006)
  [arXiv:hep-ph/0510418].

\bibitem{Pukhov:1999gg}
  A.~Pukhov {\it et al.},
  ``CompHEP: A package for evaluation of Feynman diagrams and integration  over
  multi-particle phase space. User's manual for version 33,''
  arXiv:hep-ph/9908288.

\bibitem{Pukhov:2004ca}
  A.~Pukhov,
  ``CalcHEP 3.2: MSSM, structure functions, event generation, batchs, and
  generation of matrix elements for other packages,''
  arXiv:hep-ph/0412191.

\bibitem{web} 
Our {\tt CalcHEP} files are available at 
http//theory.fnal.gov/people/kckong/6D.

\bibitem{Stelzer:1994ta}
  T.~Stelzer and W.~F.~Long,
  ``Automatic generation of tree level helicity amplitudes,''
  Comput.\ Phys.\ Commun.\  {\bf 81}, 357 (1994)
  [arXiv:hep-ph/9401258].

\bibitem{Maltoni:2002qb}
  F.~Maltoni and T.~Stelzer,
  ``MadEvent: Automatic event generation with MadGraph,''
  JHEP {\bf 0302}, 027 (2003)
  [arXiv:hep-ph/0208156].

\bibitem{Pumplin:2002vw}
  J.~Pumplin, D.~R.~Stump, J.~Huston, H.~L.~Lai, P.~Nadolsky and W.~K.~Tung,
  ``New generation of parton distributions with uncertainties from global  QCD
  analysis,''
  JHEP {\bf 0207}, 012 (2002)
  [arXiv:hep-ph/0201195].

\bibitem{CDFnote}
 CDF Collaboration, ``Combined limit for the trileptons analyses'', CDF Note 8653.

\bibitem{D0note}
  D0 Collaboration, ``Search for the Associated Production of Chargino and Neutralino in 
Final States with Two Electrons and an Additional Lepton'', D0 Note 5127-Conf.

\bibitem{Abulencia:2007zi}
  A.~Abulencia,  ``Search for new physics in lepton + photon + X events with 929-pb$^{-1}$ 
  of $p\bar{p}$ collisions at $\sqrt{s} = 1.96$ TeV'', \ arXiv:hep-ex/0702029.

\bibitem{Azatov:2007fa}
  A.~T.~Azatov,
  ``Radiative corrections to the lightest KK states in the $T2/(Z_2\times
  Z_2')$ orbifold,''
  arXiv:hep-ph/0703157.

\bibitem{Mohapatra:2002ug}
  R.~N.~Mohapatra and A.~Perez-Lorenzana,
  ``Neutrino mass, proton decay and dark matter in TeV scale universal extra
  dimension models,''
  Phys.\ Rev.\  D {\bf 67}, 075015 (2003)
  [arXiv:hep-ph/0212254].\\
  K.~Hsieh, R.~N.~Mohapatra and S.~Nasri,
  ``Dark matter in universal extra dimension models: Kaluza-Klein photon and
  right-handed neutrino admixture,''
  Phys.\ Rev.\  D {\bf 74}, 066004 (2006)
  [arXiv:hep-ph/0604154].




\bibitem{Dennis:2007tv}
  C.~Dennis, M.~K.~Unel, G.~Servant and J.~Tseng,
  ``Multi-W events at LHC from a warped extra dimension with custodial
  symmetry,''
  arXiv:hep-ph/0701158.

\bibitem{Cembranos:2006gt}
  J.~A.~R.~Cembranos, J.~L.~Feng and L.~E.~Strigari,
  ``Exotic collider signals from the complete phase diagram of minimal
  universal extra dimensions,''
  Phys.\ Rev.\  D {\bf 75}, 036004 (2007)
  [arXiv:hep-ph/0612157].

\bibitem{next}
  B.A. Dobrescu, D. Hooper, K. Kong, R. Mahbubani, work in progress.


 \end{thebibliography}
\end{document}